\documentclass[twocolumn,tighten]{aastex631}
\usepackage{CJKutf8}
\received{November 18, 2022}
\revised{June 10, 2023}
\accepted{July 20, 2023}

\usepackage{graphics}

\shorttitle{Chemical Inhomogeneities and Evolution of J1044+0353}
\shortauthors{PENG et al.}
\graphicspath{{./}{figures/}}

\begin{document}
\begin{CJK*}{UTF8}{gbsn}

\title{Using KCWI to Explore the Chemical Inhomogeneities and Evolution of J1044+0353}

\author[0000-0003-3467-6810]{Zixuan Peng (彭子轩)}
\affiliation{University of California, Santa Barbara \\
Department of Physics, Broida Hall, University of California at Santa Barbara, Santa Barbara, CA 93106, USA}
\correspondingauthor{Zixuan Peng (彭子轩)}
\email{zixuanpeng@ucsb.edu}

\author[0000-0001-9189-7818]{Crystal L. Martin}
\affiliation{University of California, Santa Barbara \\
Department of Physics, Broida Hall, University of California at Santa Barbara, Santa Barbara, CA 93106, USA}


\author[0000-0002-3867-3927]{Pierre Thibodeaux}
\affiliation{University of California, Santa Barbara \\
Department of Physics, Broida Hall, University of California at Santa Barbara, Santa Barbara, CA 93106, USA}

\author[0000-0002-0971-9535]{Jichen Zhang (张及晨)}
\affiliation{University of California, Santa Barbara \\
Department of Physics, Broida Hall, University of California at Santa Barbara, Santa Barbara, CA 93106, USA}

\author[0000-0003-3424-3230]{Weida Hu}
\affiliation{University of California, Santa Barbara \\
Department of Physics, Broida Hall, University of California at Santa Barbara, Santa Barbara, CA 93106, USA}

\author[0000-0001-8237-1515]{Yuan Li (李远)}
\affiliation{University of California, Santa Barbara \\
Department of Physics, Broida Hall, University of California at Santa Barbara, Santa Barbara, CA 93106, USA}




\begin{abstract}
J1044+0353 is considered a local analog of the young galaxies that ionized the intergalactic medium at high-redshift due to its low mass, low metallicity, high specific star formation rate, and strong high-ionization emission lines. 
We use integral field spectroscopy to trace the propagation of the starburst across this small galaxy using Balmer emission- and absorption-line equivalent widths and find a post-starburst population ($\sim 15-20$ Myr) roughly one kpc east of the much younger, compact starburst ($\sim 3-4$ Myr). 
Using the direct electron temperature method to map the O/H abundance ratio, we find similar metallicity (1 to 3 sigma) between the starburst and post-starburst regions but with a significant dispersion of about 0.3 dex within the latter.
We also map the Doppler shift and width of the strong emission lines. Over scales several times the size of the galaxy, we discover a velocity gradient parallel to the galaxy's minor axis. The steepest gradients ($\sim 30 \ \mathrm{km \ s^{-1} \ kpc^{-1}}$) appear to emanate from the oldest stellar association. We identify the velocity gradient as an outflow viewed edge-on based on the increased line width and skew in a biconical region. We discuss how this outflow and the gas inflow necessary to trigger the starburst affect the chemical evolution of J1044+0353.
We conclude that the stellar associations driving the galactic outflow are spatially offset from the youngest association, and a chemical evolution model with a metal-enriched wind requires a more realistic inflow rate than a homogeneous chemical evolution model.

\end{abstract}


\section{Introduction} \label{sec:intro}
\end{CJK*}

Gas-phase metallicity is a product of the intricate interplay among  inflows, outflows, and star formation \citep[e.g.,][]{dave_2011}. Infalling primordial gas can dilute metallicity and trigger more intense star formation. In contrast, galactic-scale outflows of gas (or galactic winds) can carry supernovae ejecta and redistribute metals to large galactic radii. The recent balance between inflows and outflows regulates the amount of gas that can be converted into stars, hence the metallicity. The intertwined processes between inflows, outflows, and star formation activities impede our ability to identify the dominant physical mechanism that affects the metal content in star-forming galaxies.

The spatial distribution of metals has been utilized to decipher the chemical evolution of galaxies. For example, negative metallicity gradients are commonly observed in spiral and disk galaxies, where the inside-out growth \citep{boissier_chemo-spectrophotometric_2000,Molla_2005,pilkington_metallicity_2012} produces the observed gradients. In contrast, the majority of dwarf irregular galaxies appear to be chemically well-mixed based on observational data \citep[e.g.,][]{KS_1997,Haurberg_2013}.

Gas outflows and inflows can impact the spatial distribution of metals within a dwarf galaxy. Several studies argue that metal-enriched galactic winds redistribute metals more effectively the shallower the gravitational potential well
\citep{ma_why_2017,tissera_oxygen_2019}. 
The metallicity scatter seen within high-redshift galaxies may reflect the chaotic accretion  processes during formation \citep{MM_2019}. The accretion of cold and metal-poor gas can produce positive metallicity gradients in star-forming galaxies \citep[e.g.,][]{cresci_gas_2010}. Interactions between galaxies are another mechanism that drives cold gas inward via gravitational torques \citep{BH_91, BH_96}, diluting central metallicities \citep[e.g.,][]{hwang_anomalously_2019,luo_evidence_2021}.

Chemical inhomogeneities in local dwarf galaxies remain a subject of ongoing debate but may identify rapidly growing low-mass galaxies \citep{Bresolin_2019,James_2020}. Consequently, a further investigation of spatial metal distributions throughout star-forming dwarf galaxies can contribute to our understanding of the interplay among star formation, inflows, and outflows.

This paper aims to measure the distribution of metals and the gas kinematics in a local analog of a reionization-era galaxy to discern between supernova-driven outflows and inflows induced by interactions or mergers. Only a small fraction of local galaxies possess nebulae with physical conditions akin to those in young galaxies observed with JWST; however, ultraviolet and optical spectroscopy effectively identify them by their emission-line strengths and ratios \citep{berg_cos_2022}. In this work, we present our analysis of the metallicity and gas kinematics of J1044+0353 based on a Keck Cosmic Web Imager (KCWI) observation (Martin et al., in prep). Similar to most local analogs, J1044+0353 has an exceptionally high [O\,{\footnotesize III}] $\lambda 5007$ equivalent width ($\sim$ 1400\AA) and a compact, UV-bright starburst. Its high specific star formation rate (log(sSFR/$\mathrm{year^{-1}}$) = $-7.73^{+0.08}_{-0.41}$\footnote{This value is adopted from the SDSS MPA-JHU DR8 catalog. This data catalog is available from \url{https://www.sdss.org/dr12/spectro/galaxy_mpajhu/}.}) and significant high-ionization emission lines (e.g., He\,{\footnotesize II} $\lambda 1640$ and C\,{\footnotesize IV} $\lambda\lambda 1548, 1550$) identify it as a young, rapidly growing galaxy  \citep{Olivier_2022}.

The structure of this paper is summarized as follows: We present the KCWI data, outline the discovery of three post-starburst stellar associations, and detail the line-fitting measurements in Section \ref{sec:Data_Reduce_line_fitting}. In Section \ref{sec:results}, we estimate the ages of the stellar populations within these stellar associations. We then measure the ``direct'' oxygen abundance for each post-starburst association and the starburst region. To visualize the kinematics and metallicity distribution of this galaxy, we apply an adaptive binning algorithm to generate a peculiar velocity map and a metallicity map. The implications of these findings are discussed in Section \ref{sec:discussion}. We summarize our key results in Section \ref{sec:conclusion}. Throughout this paper, we use $\mathrm{Z_{\odot}} = 0.0134$ for solar metallicity, corresponding to 12 + log(O/H) = 8.69 \citep{Asplund_2021}.  

\vspace{+1em}

\section{Data Processing and Line Fitting Model} \label{sec:Data_Reduce_line_fitting}

\begin{figure*}
\centering
\includegraphics[width=1\textwidth]{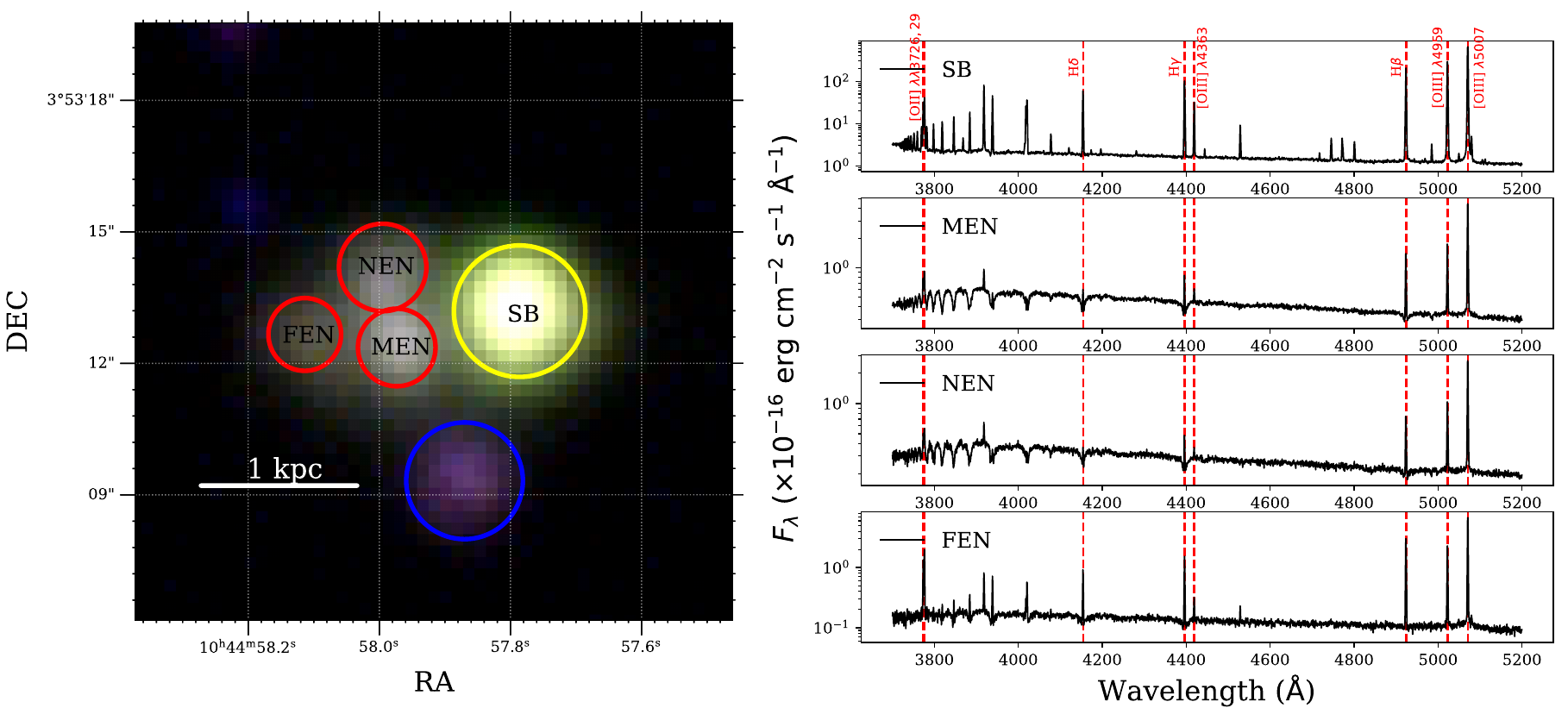}
\caption{{\em Left:} Color composite image of J1044+0353 constructed from the DECalS g,r,z-images. The stretch is an inverse hyperbolic sine function. The yellow circle (SB) is the Sloan-like circular aperture with a radius of 1\farcs5 covering this galaxy's starburst region. The three red circles (FEN, MEN, and NEN) identify lower surface brightness stellar associations. The furthest is  4\farcs9 (1.4 kpc) east of the starburst. 
The blue circle represents the background galaxy identified by Martin et al. (in prep). {\em Right:} The extracted spectrum of each stellar association. \label{fig:j1044_rx10}} 
\end{figure*}

\subsection{Target}\label{subsec:target}
\defcitealias{berg_carbon_2016}{B16}
\defcitealias{berg_cos_2022}{B22}
The corrected luminosity distance of J1044+0353 is $59$ Mpc from the peculiar velocity model of \citet{willick_homogeneous_1997}, but another model gives a distance of $55$ Mpc \citep{masters_2005}. We carry this systematic uncertainty forward throughout our analysis. Masses and SFRs adopted from the published literature are converted to the adopted distance ($57 \pm 2$ Mpc) in Table \ref{table:oh_physical_literature}. Both this study and \citet{berg_cos_2022} use the \citet{chabrier_galactic_2003} IMF, but the MPA-JHU catalog adopt the \citet{Kroupa_2001} IMF. We neglect the $\sim$ 0.03 dex systematic difference between these two IMFs, but we propagate this difference to the uncertainty of each quantity \citep{Moustakas_2013}.
The stellar mass of J1044+0353, $10^{6.83^{+0.41}_{- 0.26}} M_{\odot}$, identifies it as a dwarf galaxy. \cite{berg_carbon_2016} and \cite{berg_cos_2022} (hereafter B16 and B22) measure a very low metallicity of 12 + log(O/H) = $7.45 \pm 0.02$ and $7.44 \pm 0.01$ respectively. 

Figure \ref{fig:j1044_rx10} shows a color image constructed from the g, r, and z band DECalS images \citep{decals_16,desi_2019}.\footnote{Images can be downloaded from \url{https://www.legacysurvey.org/}} The spatial resolution, 0\farcs262 pixel$^{-1}$, is limited by atmospheric seeing and is not sufficient to resolve individual star clusters. However, several local intensity maxima are detected east of the compact starburst.
In Section \ref{subsubsec:oh_gradient}, we will argue that the starburst is offset from the centroid of the stellar mass.
In contrast to most dwarf galaxies, there is no obvious low surface brightness component.

\vspace{-1.0em}
\setlength{\tabcolsep}{9pt}
\begin{deluxetable*}{ccccccc}
\tablenum{1}
\tablecaption{Physical Properties of J1044+0353 Collected from the Literature}
\tablewidth{0pt}
\tablehead{
\colhead{Source} & \colhead{Aper. log($M_\ast$) ($M_\odot$)} & \colhead{Tot. log($M_\ast$) ($M_\odot$) } & \colhead{Aper. log(SFR) ($M_\odot$ $\mathrm{yr^{-1}}$)} & \colhead{Tot. log(SFR) ($M_\odot$ $\mathrm{yr^{-1}}$)} 
}
\startdata
MPA-JHU & 6.71$^{+ 0.47}_{- 0.06}$ & 6.83$^{+ 0.45}_{- 0.05}$ & -0.79$^{+ 0.05}_{- 0.05}$ & -0.82$^{+ 0.06}_{- 0.06}$   \\
\citetalias{berg_cos_2022} & 6.16$^{+ 0.23}_{- 0.09}$ & 6.83$^{+ 0.41}_{- 0.26}$ & -0.87$^{+ 0.08}_{- 0.11}$ & -0.56$^{+ 0.11}_{- 0.14}$  \\
\enddata
\tablecomments{Stellar mass and SFR (aperture and total) of J1044+0353 extracted from the MPA-JHU Catalog (Row 1) and Table 6 of \citetalias{berg_cos_2022} (Row 2) are converted to the \citet{chabrier_galactic_2003} IMF and the adopted distance ($57 \pm 2$ Mpc) in this work. Columns 2 and 3 list the aperture and total stellar masses. Columns 4 and 5 list the aperture and total SFR. For row 1, the value for each column is extracted from the median estimate of the (log) stellar mass or SFR PDF. The 1-$\sigma$ error of each value is calculated from the 16th and 84th percentiles of the (log) stellar mass or SFR PDF. For row 2, the stellar masses or SFRs are derived using BEAGLE SED fitting for both the light within the COS aperture and the entire galaxy, respectively.}
\label{table:oh_physical_literature}
\end{deluxetable*}
\vspace{-1.0em}

\subsection{Data Cube} \label{subsec:data_cube}

KCWI is a sensitive optical (350 - 560 nm) integral field spectrograph on the Keck II telescope
\citep{morrissey_keck_2018}. 
The KCWI observations of J1044+0353 (R.A.=10:44:57.80, Decl.=+03:53:13.2) were obtained on 2018 January 14 with an average seeing of 0\farcs9. We reduced raw frames and built data cubes using
the KCWI Data Extraction and Reduction Pipeline v.1.1.0 \citep[][KDERP]{morrissey_keck_2018}. We employed
the CWITools Python package\footnote{https://github.com/dbosul/cwitools} \citet{OC_2020} to coadd the individual data cubes, creating a mosaic cube with $0\farcs146 \times 0\farcs146 \times 0.5 \text{\AA}$ voxels.
We refer the reader to our companion paper (Martin et al., in prep) for details about the observations and data reduction. Here we summarize the properties of the data cube central to the metallicity and kinematic measurements presented in this paper.

The data cube is a mosaic of three pointings of the 8\arcsec\ by 20\arcsec\ field of view of the small slicer. The observations were made with the BL grating, which provides contiguous wavelength coverage from the atmospheric limit to the dichroic cutoff at 5600 \AA. In this configuration, the slits of the image slicer are 0\farcs34 wide, and the average resolution over the bandpass is $R \approx 3600$. Three 1200~s and three 300~s integrations were combined to form the full mosaic. In the 1200~s frames, the cores of the [O\,{\footnotesize III}] $\lambda\lambda 4959, 5007$ and $\mathrm{H\beta}$ emission lines saturate in a few spaxels centered on the compact starburst. We measure the [O\,{\footnotesize III}] $\lambda 4959$ and $\mathrm{H\beta}$ line fluxes directly from a mosaic of the 300~s frames in these regions. The spectral resolution $R \approx 3600$ produces
an instrumental line width of about 83 km/s FWHM.
The atmospheric seeing was estimated to be 0\farcs90, and a relative flux calibration was performed using the standard star Feige 34.

\subsection{Stellar Associations} \label{subsec:source_detection}

We utilized {\tt\string Photutils} \citep{larry_bradley_2022_6825092} to carry out an automatic identification of objects on the raw r-band DECalS image of J1044+0353, where each object is a group of unresolved stellar clusters that we called a stellar association hereafter. 
To better trace the boundary of each stellar association, we generated a 2D background model using an instance of {\tt\string MedianBackground} (i.e, the sigma-clipped median) in {\tt\string Photutils}.
Before the deblending process, the constructed 2D background model was subtracted from the r-band image.
To deblend the local intensity maxima detected east of the starburst, we established the threshold for identifying sources at 10$\sigma$, with a minimum of one pixel above this threshold required for analysis. We found that only this minimum threshold value can deblend the local maxima into three stellar associations found by eye in Figure \ref{fig:j1044_rx10}. We also employed 1024 deblending levels and a deblending contrast of 1\%. Combining with the starburst, we refer to four stellar associations as the far-east nucleus (FEN),  north-east nucleus (NEN),  middle-east nucleus (MEN), and compact starburst (SB) (shown in Figure \ref{fig:j1044_rx10}). We will show that these eastern associations are significantly older than SB (Section \ref{subsec:ifu_structure}), so we refer
to their collection as the post-starburst region (PSB).

Following the deblending map, we utilized {\tt\string Petrofit} \citep{Geda_2022} to calculate the Petrosian radius and flux \citep{Petrosian_1976} for each stellar association. We used the half-light radius, $r_{50}$, to define the size of each stellar association. An accurate determination of $r_{50}$ is related to the radius that encloses the total flux, $r_{\mathrm{total}}$. To attain $r_{\mathrm{total}}$, we must multiply the Petrosian radius with a constant $\epsilon$. $\epsilon$ is set to 2 by default in {\tt\string Petrofit}, which is appropriate for a Sérsic light profile with the Sérsic index $n = 1$ \citep[i.e., an exponential light profile;][]{Yasuda_01}.
To find the best-fitting $\epsilon$ of each detected stellar association, we constructed a grid of ideal Sérsic profiles, convolved with the DECalS coadded PSF (point spread function) of J1044+0353\footnote{\url{https://www.legacysurvey.org/viewer/data-for-radec/?ra=161.2415&dec=3.8870&layer=ls-dr9&ralo=161.2212&rahi=161.2619&declo=3.8765&dechi=3.8970}}, with different half-light radii and Sérsic indices to implement the Petrosian correction\footnote{\url{https://petrofit.readthedocs.io/en/latest/correction_grids.html}}. After obtaining the corrected $r_{50}$ for each object, we adopted a common practice in the literature \citep[e.g.,][]{Law_2012}, to convert $r_{50}$, calculated along the semimajor axis, to a circularized effective radius $r_c = r_{50} \sqrt{b/a}$ (b and a are the minor and major axis, respectively), where the area enclosed by these two radii are the same (see Table \ref{table:photometry_eastern}).

\vspace{-1em}
\setlength{\tabcolsep}{12pt}
\begin{deluxetable*}{cccccc}
\tablenum{2}
\tablecaption{WCS Coordinates and Petrosian Properties of the Eastern Stellar Associations.}
\tablewidth{0pt}
\tablehead{
\colhead{} & R.A. (J2000) & Decl. (J2000) & $r_{50}$ (arcsec)\tablenotemark{a} & $r_{c}$ (arcsec)\tablenotemark{b} & Ellipticity\tablenotemark{c}}
\startdata
MEN & 10:44:57.9734  & +3:53:12.362 & 0.942 & 0.886 & 0.115 \\
NEN & 10:44:57.9948  & +3:53:14.187 & 1.247 & 0.998 & 0.359 \\
FEN & 10:44:58.1137  & +3:53:12.660 & 0.932 & 0.830 & 0.207 \\
\enddata
\tablenotetext{a}{Half-light radius measured as the Petrosian half-light radius.}
\tablenotetext{b}{The circularized effective radius is calculated as $r_c = r_{50} \sqrt{b/a}$, where $b$ and $a$ represent the minor and major axes, respectively.}
\tablenotetext{c}{Ellipticity = $1 - \frac{b}{a}$, where $b/a$ is the minor/major axis ratio.}
\label{table:photometry_eastern}
\end{deluxetable*}
\vspace{-1em}

\subsection{Extraction of Spectra} \label{subsec:extraction_spectra}
Using {\tt\string pyregion}\footnote{\url{https://github.com/astropy/pyregion}}, we extracted optical spectra of the stellar associations circled in the left panel of Figure \ref{fig:j1044_rx10} to estimate their stellar population ages, stellar mass, and gas-phase oxygen abundance. 
The right-hand panel of Figure \ref{fig:j1044_rx10} shows the emission lines of the starburst region are two orders of magnitude stronger than those of the three eastern associations. 

We scaled the error spectra because the variance produced by the KCWI Data Reduction Pipeline underestimates the noise
\citep{osullivan_flashes_2020}.
To determine the variance rescaling factor, we compare the median variance with the square of the root-mean-square noise of the continuum near strong emission lines. The typical scaling factor is $\sim 2.0$ for all extracted spectra.

For purposes of mapping the spatial variations in peculiar velocity and metallicity, we can measure these physical properties at smaller scales.
The data cube comprises 15,456 spaxels; however, the signal-to-noise ratio (SNR) of numerous emission lines extracted from individual spaxels is insufficient for providing robust constraints on physical properties.
%

Facing this issue, we applied an adaptive binning algorithm\footnote{\url{https://github.com/pierrethx/MVT-binning}} based on the Centroidal Voronoi Tessellation (CVT) method described in \cite{cappellari_adaptive_2003} that improves the SNR of the emission lines by aggregating spaxels, giving us a stronger constraint on the local physical properties.
Since the size of each spaxel is $(0\farcs146)^2$, we have to make sure the minimum bin has $\pi (3 \ \mathrm{spaxels})^2 \approx 28$ spaxels in total, the resolution limit. We require this minimum bin size during the initial formation of the bins.

Our adaptive binning code runs on emission-line images. The local continuum has been fitted and subtracted at each spaxel. Random noise fluctuations cause half the spaxels with no (or low) emission-line intensity to have negative fluxes. The original CVT method cannot accommodate negative spaxels, so we devised a modification that initially masks the negative spaxels during bin formation before applying the binning to the original data. 

This modified CVT method has difficulty maintaining minimum bin size. Thus, the version we use (method ``VT" in the code) lacks the centroidal feedback that equalizes SNR between bins by crowding them in high SNR regions. This ``VT" method prioritizes the size of the bins over the scatter of bin SNRs around the target SNR.

\subsection{Line-Fitting Model} \label{subsec:line-fitting}
We used the fitting algorithm based on the Levenberg-Marquardt method from the package {\tt\string LMFIT}
\citep{lmfit}. Each iteration of this algorithm attempts to improve the fit by varying the parameters along the gradient of improvement in the multi-dimensional parameter space. 

Each extracted spectrum is continuum-subtracted before fitting any line profile. To estimate the local continuum level around each line, we first extract two wavelength regions on both sides of this line and fit them with a polynomial of the first order. We mask out other strong emission or absorption lines around the line profile and use a sigma-clipping algorithm \citep{cont_2018} to ensure an accurate local continuum determination. Then, we extrapolate this first-order polynomial to estimate the continuum level within the line profile. 

A single-Gaussian model suffices for fitting symmetric emission line profiles without a second velocity component. We find two exceptions: (1) the Balmer lines with underlying, stellar absorption troughs and (2) emissions lines (e.g., [O\,{\footnotesize III}] $\lambda 4959$) with two velocity components, sometimes making the combined profile asymmetric. We conduct an F-test on each exception at a 3-sigma significance level to determine whether a two-component model (see Section \ref{subsubsec:2comp_em_ab} and Section \ref{subsubsec:2comp_em}) performs statistically better than the single-Gaussian model \citep[]{xu_classy_2022}. 

We constrain the gas kinematics by assuming the velocity center and width of each component (stellar or nebular) in the intended emission lines (including [O\,{\footnotesize II}] $\lambda\lambda 3726,29$, [O\,{\footnotesize III}] $\lambda 4363$, [O\,{\footnotesize III}] $\lambda 4959$, H$\delta$, H$\gamma$, and H$\beta$) are the same.
Here we summarize the details of each two-component fitting strategy.



\subsubsection{Two-component Fitting: Emission + Absorption} \label{subsubsec:2comp_em_ab}

For Balmer lines with significant stellar absorptions, we must be aware of their impacts on the observed line ratios. We prefer a ``Gaussian plus Lorentzian'' model (GL model) to fit both the emission and absorption components simultaneously (Figure \ref{fig:extinction_maps}). The Gaussian (or Lorentzian) function of the GL model is used to fit the emission (or absorption) component. All three eastern stellar associations have obvious stellar absorption troughs that require a GL model to fit their Balmer lines well. 


\subsubsection{Two-component Fitting: Narrow+Broad Emissions} \label{subsubsec:2comp_em}

\begin{figure*}[tb]
\gridline{\fig{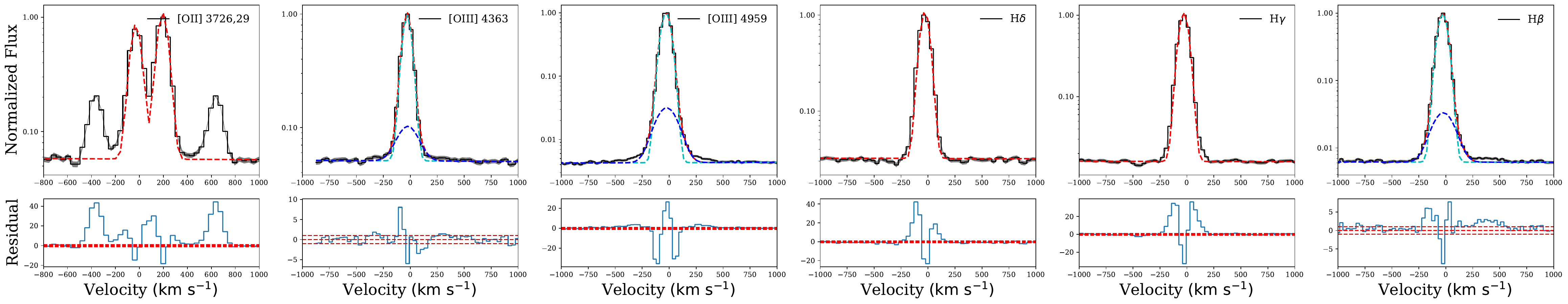}
{1.00\textwidth}{\textbf{{\large (SB)}}}}
\gridline{\fig{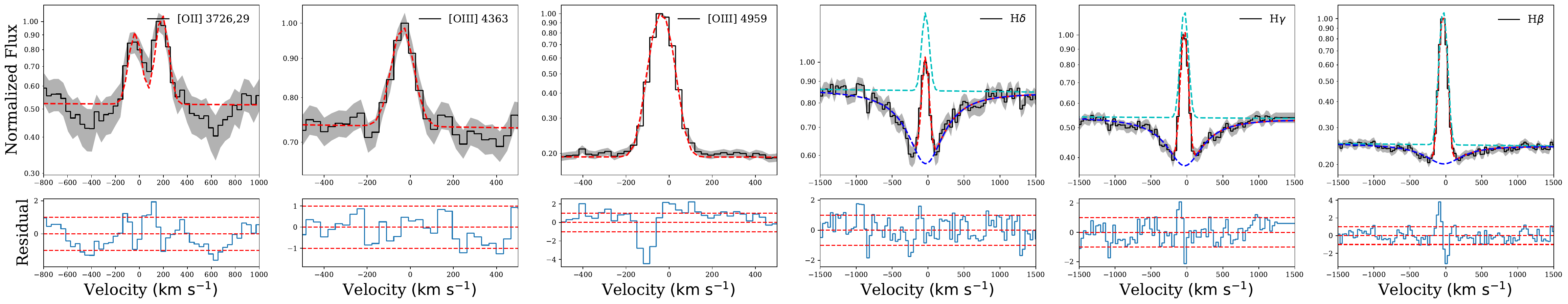}
{1.00\textwidth}{\textbf{{\large (MEN)}}}}
\gridline{\fig{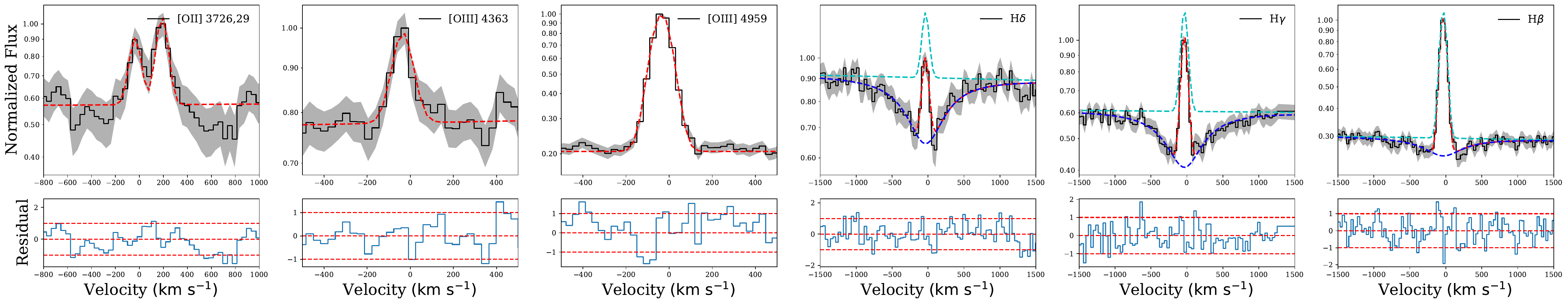}
{1.00\textwidth}{\textbf{{\large (NEN)}}}}
\gridline{\fig{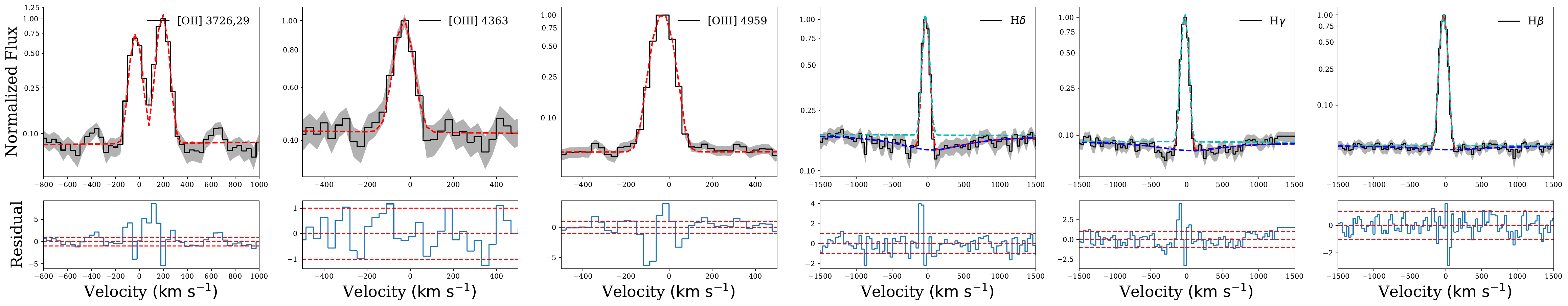}
{1.00\textwidth}{\textbf{{\large (FEN)}}}}
\caption{Fittings of Balmer lines $\mathrm{H\beta}$, $\mathrm{H\gamma}$, and $\mathrm{H\delta}$ for each stellar association labeled in Figure \ref{fig:j1044_rx10}. For each stellar association, the best-fitting model (red dashed line) is plotted with the observed line profile (black solid line) in the first row. For double-Gaussian or Gaussian-Lorentzian fittings, the broad and narrow components are shown as the blue dashed line and the cyan dashed line, respectively. The grey areas represent the 1-$\sigma$ uncertainties of raw data. The residual (i.e., (data - model)/error) plot is included in the second row. The two red dashed lines in the residual plot represent residual = $\pm 1$. \label{fig:extinction_maps}}
\end{figure*}

Apart from significant stellar absorptions, we observe that $H\beta$, [O\,{\footnotesize III}] $\lambda 4363$, and [O\,{\footnotesize III}] $\lambda 4959$ lines in the starburst region (shown in Figure \ref{fig:extinction_maps}) or the [O\,{\footnotesize III}] $\lambda 5007$ lines around the post-starburst region (shown in Figure \ref{fig:asymmetric_line_profiles}) exhibit another component that is poorly fitted by single-Gaussian models. To improve the fit, we introduce an additional Gaussian function with a positive amplitude and increased width. Nonetheless, substantial residuals persist in the broad wings of double-Gaussian fits (e.g., the starburst's [O\,{\footnotesize III}] $\lambda$4959 line in Figure \ref{fig:extinction_maps}), implying the presence of a ``very-broad'' velocity component (FWHM $\sim 700 \ \mathrm{km \ s^{-1}}$; Martin et al., in prep) or the non-Gaussian nature of these broad wings \citep[e.g., follows a velocity field of power law in][]{Carr_2022}. However, these residuals constitute $\lesssim 1 \%$ of the total flux, indicating a negligible effect on electron temperature and oxygen abundance measurements. 

Utilizing the best-fitting parameter values, we compute the flux of each component in every line profile. For lines that require a double-Gaussian fit, we report in Table \ref{table:j1044_physical} the flux of the narrow and broad components, respectively. In this study, we interpret the narrow component as gas tracing virial motions within the examined region and the broad component as turbulent outflowing gas. Therefore, we rely solely on the narrow component's flux to determine the nebular physical properties of each region (except intrinsic $A_V$ measurement in SB; see Section \ref{subsubsec:extinction_sb}). 

The flux uncertainty is obtained from the {\tt\string LMFIT} covariance matrix. 
To determine the statistical errors for line ratios, we follow a Monte Carlo approach to perturb each line profile according to its error spectrum, recalculate the corresponding line ratio, and repeat the previous steps 1,000 times to build up a distribution of perturbed line ratios. Then, the 68 percentile width of the distribution accounts for the statistical error of this line ratio.

\begin{subsection}{Extinction} \label{subsec:extinction}
In this work, we use the Galactic extinction curve interpolated from Table 3 of \cite{Fitzpatrick_1999} to correct for the influences from the Milky Way dust. We adopt the mean $E(B-V)$ value ($0.03670$) from the \textit{NASA/IPAC Infrared Science Archive}\footnote{\url{https://irsa.ipac.caltech.edu/applications/DUST/}} \citep{schlafly_measuring_2011}. For intrinsic extinction, we use the SMC extinction curve interpolated from Table 4 (SMC Bar) of \cite{gordon_2003}.



The intrinsic extinction $A_V$ is determined from the Balmer decrement. Reddening attenuates the bluer line more than the redder line, so the observed Balmer ratio should always be larger than the intrinsic ratio. However, in regions of J1044+0353 with low dust content, minimal extinction corrections can occasionally result in unphysical extinction values.


Here we summarize our method (Section \ref{subsubsec:intrinsic_balmer_ratio}) to determine the intrinsic $A_V$ of each stellar association (Section \ref{subsubsec:extinction_sb} and \ref{subsubsec:extinction_psb}) and our approach to dealing with unphysical extinction values (Section \ref{subsubsec:extinction_oh_map}) when we create the metallicity map.

\subsubsection{Intrinsic Balmer Ratio} \label{subsubsec:intrinsic_balmer_ratio}
We need to know the intrinsic Balmer line ratios well to obtain physical reddening values. In this study, we extract the intrinsic ratios from \cite{storey_recombination_1995}. Compared to $\mathrm{n_e}$, the intrinsic ratios depend more on $\mathrm{T_e}$ and the assumption of the H\,{\footnotesize II} region being optically thin or thick (Case A or Case B). Case A and B's intrinsic ratio is higher for the H\,{\footnotesize II} region with a higher $\mathrm{T_e}$. At fixed $\mathrm{T_e}$ and $\mathrm{n_e}$, the intrinsic ratios of Case A regions are smaller than the ones of Case B regions. The stellar-mass-weighted average $\mathrm{T_e}$ and $\mathrm{n_e}$ for these four stellar associations are  $19727^{+347}_{-311}$ K and $\mathrm{99^{+57}_{-37} \ cm^{-3}}$, respectively. Hence, we apply the Balmer line intrinsic ratios at $\mathrm{T_e} = 20,000$ K and $\mathrm{n_e} = \mathrm{100 \ cm^{-3}}$  (for Case B (or Case A), we have $\mathrm{\mathrm{H\beta} / \mathrm{H\gamma} = 2.10}$ (or 2.07) and $\mathrm{\mathrm{H\gamma} / \mathrm{H\delta} = 1.80}$ (or 1.78)) to determine $A_V$. 

The $A_V$ values derived from Balmer line ratios H$\beta$/H$\gamma$ and H$\gamma$/H$\delta$ typically differ. We adopt the $A_V$ determined from the line ratio of H$\beta$ and H$\gamma$ lines, less sensitive to stellar absorption, as a more accurate estimate of the true value.



\subsubsection{Extinction of the Starburst Region} \label{subsubsec:extinction_sb}
Due to the broad emission wings of $\mathrm{H}\beta$ line, we use a double-Gaussian model to fit the broad and narrow components simultaneously (Section \ref{subsubsec:2comp_em}). The contribution of the broad component to the total $\mathrm{H}\beta$ flux is $\sim 6\%$. In contrast, the low SNR of the broad components in $\mathrm{H}\gamma$ and $\mathrm{H\delta}$ lines prevents using a double-Gaussian model, which did not significantly improve the fitting as it could not pass the F-test. Moreover, the contribution of the broad component in $\mathrm{H}\gamma$ and $\mathrm{H\delta}$ lines are negligible (less than $\sim 3 \%$), so a single-Gaussian model is used instead. Due to the unresolved broad component in H$\gamma$, we utilize the total line flux ratio of H$\beta$ and H$\gamma$ lines to determine the starburst's $A_V$.


Our derived $A_V$ ($0.20^{+0.10}_{-0.10}$) of the starburst region is similar to the one reported by \citet{berg_characterizing_2021} \defcitealias{berg_characterizing_2021}{B21} (hereafter B21; $A_V = 0.24^{+ 0.03}_{- 0.02}$) but smaller than that determined by \citetalias{berg_carbon_2016} ($A_V = 0.37^{+ 0.03}_{- 0.03}$), where they do not correct for Milky Way extinction. 
Both studies use the relative intensities of the four strongest Balmer lines ($\mathrm{H\alpha / H\beta}$, $\mathrm{H\gamma / H\beta}$, $\mathrm{H\delta / H\beta}$) to determine the intrinsic dust reddening value. This value is then used to reddening-correct other optical emission lines using the \cite{cardelli_1989} extinction law. Our observed Balmer line ratio $\mathrm{H\beta} / \mathrm{H\gamma} = 2.21 \pm 0.07$ is consistent with \citetalias{berg_characterizing_2021} ($2.19 \pm 0.02$). These systematic differences cause minor variations of $A_V$ in the starburst region between our study and \citetalias{berg_characterizing_2021}. 


\subsubsection{Extinction of the Post-Starburst Region} \label{subsubsec:extinction_psb}
FEN ($A_V = 0.20^{+0.04}_{-0.04}$) and NEN ($A_V = 0.21^{+0.13}_{-0.12}$) have similar intrinsic reddening values as the SB.  
On the contrary, we find the correction factors for intrinsic extinction are large in MEN ($A_V = 0.56^{+0.09}_{-0.09}$). This result is unexpected in a low-metallicity environment because the color
excess, $E(B-V)$, typically increases with metallicity \citep[e.g.,][]{Zahid_2014_dust}. 
If confirmed with $\mathrm{H\alpha / H\beta}$ measurements, the implied dust lane would run roughly north-south between the SB and FEN. Similar dust lanes have been seen after galaxy mergers \citep{Whitmore_2010}. 

\subsubsection{Regions Where Case~B Fails} \label{subsubsec:extinction_oh_map}
Some bins show negative intrinsic $A_V$ when we create the metallicity map (the binning criteria detailed in Section \ref{subsubsec:oh_gradient}).
To deal with this issue, we propagate the error spectrum of each Balmer line 1000 times to get the distribution of the observed Balmer line ratio. If the intrinsic Balmer ratio falls within the 95$\%$ percentile width of the distribution, we attribute zero extinction to those bins (i.e., $A_V = 0$), given that the noise within the observed Balmer line ratios can occasionally result in an observed line ratio that is smaller than the intrinsic ratio.


A few bins show unphysical reddening values, such that the intrinsic ratio for Case B resides outside the 95$\%$ percentile width of their observed Balmer ratio distribution. Figure \ref{fig:balmer_ratio_map} shows that many of these bins are immediately north of SB. 
Since the galaxy might be optically thin in this direction, we calculated $A_V$ using the intrinsic Case A ratio and obtained physical values for most of the bins, depicted by the non-overlapping red contours in Figure \ref{fig:balmer_ratio_map}. We suggest that these optically thin bins, corresponding to holes in the H\,{\footnotesize I} distribution, likely provide an escape channel for ionizing photons.

\begin{figure}[htb]
    \centering
    \includegraphics[width=0.5\textwidth]{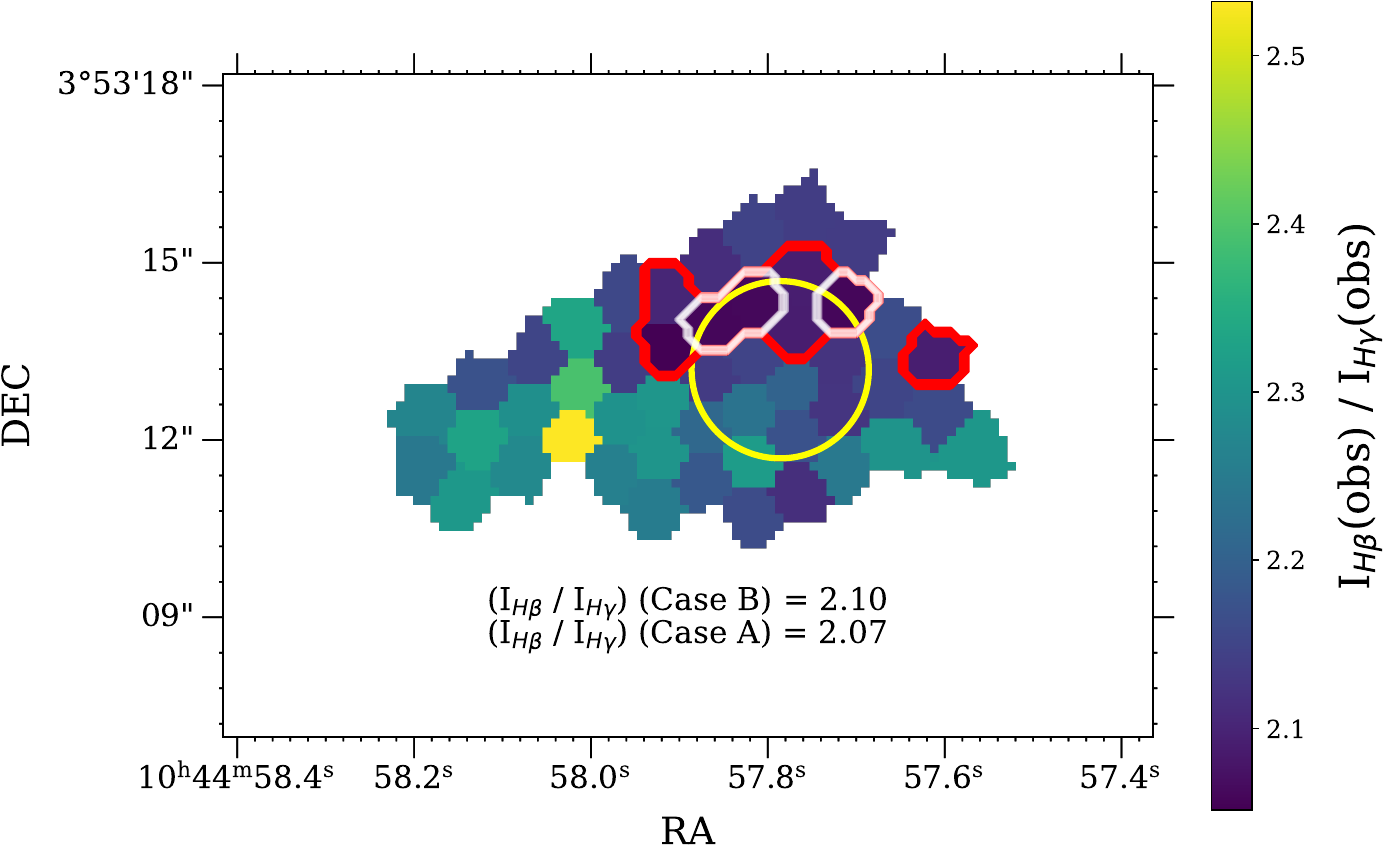}
\caption{The observed H$\beta$/H$\gamma$ line ratio map. Some regions surrounding the northern part of SB exhibit unphysical extinction values when their observed ratios are compared to the intrinsic ratio for Case B (red contours) or Case A (white contours). These optically thin regions, corresponding to holes in the H\,{\footnotesize I} distribution, might reveal the escape channels for ionizing photons.}  \label{fig:balmer_ratio_map}
\end{figure} 

\end{subsection}

\section{Results from Integral Field Spectroscopy} \label{sec:results}
In this section, we estimate the stellar population age of each stellar association (Section \ref{subsec:ifu_structure}). We also map the gas kinematics across the ionized nebula
(Section \ref{subsubsec:velocity_gradient}). 
We use the direct method to measure gas-phase oxygen abundance and examine how metallicity varies across these stellar associations (Section \ref{subsubsec:sb_metallicity} $\&$ \ref{subsubsec:older_nuclei}) and
with distance from the stellar-mass-weighted center.
(Section \ref{subsubsec:oh_gradient}).

\subsection{Star Formation History} 
\label{subsec:ifu_structure}




Previous studies focus on analyzing physical properties determined from the starburst region of J1044+0353 (e.g., \citetalias{berg_carbon_2016, berg_cos_2022}). The Balmer lines in the starburst spectrum show minimal stellar absorption, consistent with a very young stellar population. However, an inspection of the KCWI data cube east of the compact starburst revealed significant absorption troughs under Balmer lines, indicating they have older stellar populations (see Figure \ref{fig:extinction_maps}). To understand the propagation of the starburst, we measured the strength of Balmer lines in spectra of the stellar associations defined in  Figure \ref{fig:j1044_rx10}. 

\begin{figure*}[htb]
\centering
\includegraphics[width=1.0\textwidth]{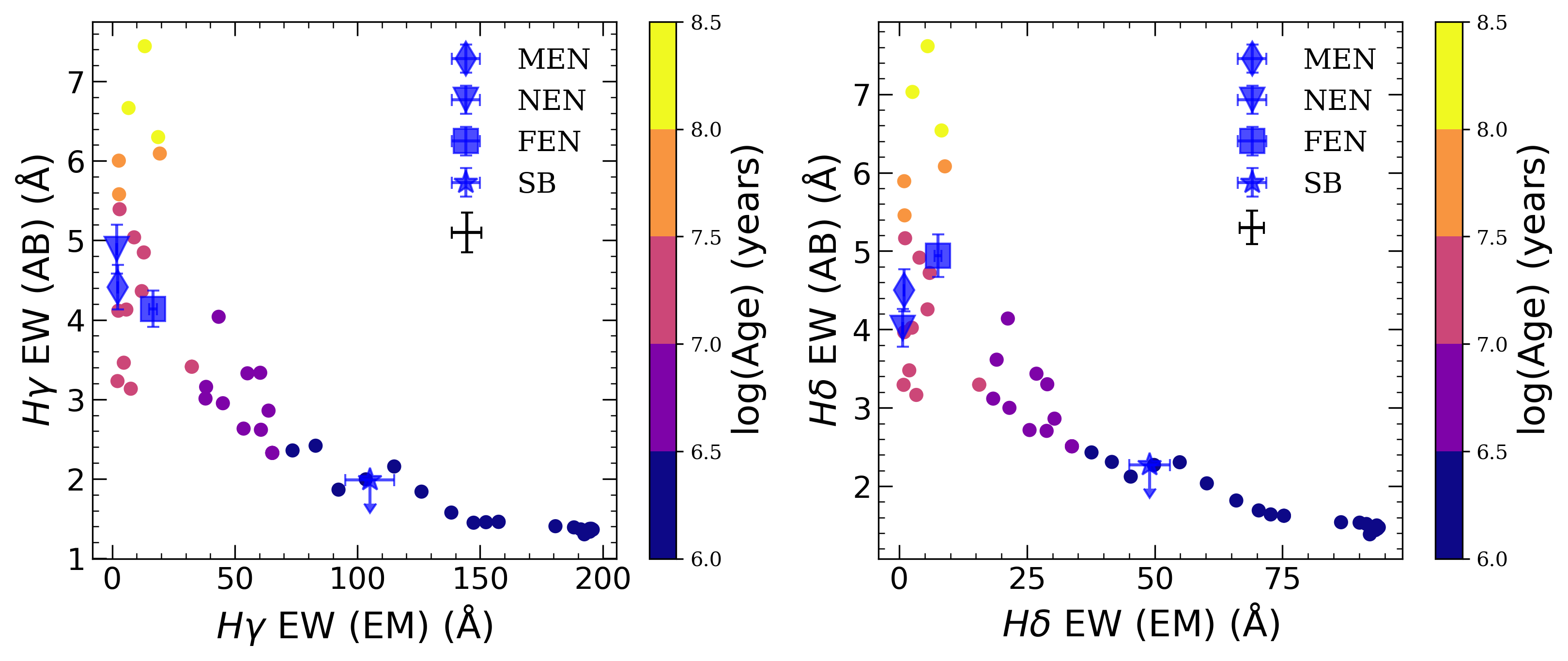}
\caption{The positions of stellar associations lie in the EW space constructed by BEAGLE's mock galaxy spectra. The composite SFH model ranges from $10^6$ to $10^{8.5}$ years. For both panels ({\em Left:} $\mathrm{H\gamma}$; {\em Right:} $\mathrm{H\delta}$), the EW of the emission portion is on the x-axis, and the EW of the absorption portion is on the y-axis. The marker for each stellar association is shown as a diamond (MEN), an inverted-triangle (NEN), a square (FEN), and a star (SB), respectively. For both panels, the absorption EWs of the starburst region cannot be accurately measured due to its weak absorption troughs. Therefore, we assume the absorption EWs of the BEAGLE mock profile at the corresponding age (i.e., $10^{6.48}$ years) to be the upper limits for those of the starburst region. The circular points are BEAGLE mock galaxies that are color-coded by their ages. Within each panel, the upper right corner displays the representative error bars of the measured emission and absorption EWs for mock galaxies.\label{fig:beagle_age}}
\end{figure*}

To assign stellar population ages, 
we use a stellar population synthesis model, BEAGLE \citep{chevallard_modelling_2016}, to build mock galaxy spectra at different ages and measure equivalent widths (EWs) for both emission (EW(EM)) and absorption (EW(AB)) portions of the Balmer line profiles. We take the absolute values of EW(EM) and EW(AB) to compare the relative strength of emission and absorption portions of each line profile (see Table \ref{table:j1044_physical}). Each EW grid has ages ranging from $10^6$ to $10^{8.5}$ years. Although the Balmer profiles are fitted simultaneously (Section \ref{subsec:line-fitting}), the absorption trough in the $\mathrm{H\beta}$ line of each eastern stellar association is too weak for precise fitting, yielding roughly one dex lower measured EW(AB) than the corresponding mock spectrum. As a result, we rely on the $\mathrm{H\gamma}$ and $\mathrm{H\delta}$ lines to construct the EW grids shown in Figure \ref{fig:beagle_age}. 

In our simulation, we assume the initial mass function of \cite{chabrier_galactic_2003} and the two-component dust attenuation model of \cite{charlot_simple_2000}. To determine an appropriate SFH that is consistent with our EW measurements in Table \ref{table:j1044_physical}, we first generate mock galaxy spectra for a constant-SFH model adopted in \citetalias{berg_cos_2022} and a pure SSP (simple stellar population) model. The constant-SFH model overestimates the absorption EWs of the older stellar associations at their emission EWs. Instead, the pure SSP model has zero emission EWs at age $\gtrsim 10^7$ years because all O stars die off. Hence, we propose a composite SFH that includes a recent burst with $\mathrm{log(SFR_c/M_{\odot} \ yr^{-1})} = -0.85$ and a timescale $\Delta t_{\mathrm{SFR_c}} = 10^6 \ \mathrm{year}$ (see Table \ref{table:beagle}), in addition to the first burst occurring at the ignition of the star formation.
If we set the $\mathrm{SFR_c}$ as the total SFR from \citetalias{berg_cos_2022} (Table \ref{table:oh_physical_literature}), the offsets in our age estimations are within the uncertainties for all regions.


BEAGLE simulates the spectrum from both stars and photoionized gas. The value of each stellar or nebular fixed parameter is listed in Table \ref{table:beagle}. We assume \textit{in situ} star formation of these stellar associations so we can input the same fixed parameters in BEAGLE for each stellar association. 

We estimate the age of each stellar association (see Table \ref{table:nucleus_age_mass}) by finding the nearest model points in the EW space (Figure \ref{fig:beagle_age}). Our age estimation for SB ($3.02 \pm 0.03$ Myr) is consistent with the finding of \cite{Olivier_2022} ($4.01 \pm 1.13$ Myr), who modeled the FUV spectrum, encompassing photospheric and stellar wind lines that are sensitive to the presence of the most massive stars. The uncertainties of NEN and MEN are overlapped with each other in both $\mathrm{H\gamma}$ and $\mathrm{H\delta}$ EW spaces, which suggests they have similar stellar population ages ($10^{7.2 \pm 0.1}$ years). Even though the stronger emission EWs of these two Balmer lines might imply that FEN's stellar population is relatively younger, the estimated age is relatively older ($10^{7.3 \pm 0.05}$ years). These age estimations are based on the argument that the stellar and the gas-phase metallicities are similar, as found in a sample of low-redshift star-forming galaxies \citep{Chisholm_2019}. Instead, if we assume the stellar metallicity is 0.2 dex lower than the gas-phase metallicity \citep{Rossi_2017}, the estimated ages of SB and FEN would stay the same. However, the measured age of MEN or NEN would increase by a factor of $\sim 1.3$ to $10^{7.33 \pm 0.06}$ years. 


The stellar mass of each stellar association can be estimated by the scaling needed to match the model continuum (at $10^{6.8} M_{\odot}$) to the observed continuum.
To avoid the regions with strong emission lines, we only use the continuum level with wavelengths ranging from 5100 to 5500 \r{A}. This method shows MEN (or FEN) has the largest (or smallest) stellar mass (Table \ref{table:nucleus_age_mass}). One main caveat of this approach is that we do not explore other complicated SFH models in determining the stellar mass. Even though the estimated starburst's stellar mass from \citetalias{berg_cos_2022} (assume constant SFH) lies within the uncertainty of our result, the MPA-JHU value lies 2$\sigma$ away from ours. Therefore, the assumption that J1044+0353 follows the composite SFH model might introduce significant systematic uncertainties and lead to deviations from previous studies. Besides this issue, since the mass-to-light ratios in the BEAGLE mock galaxies are extremely sensitive to age, an accurate age estimation is needed to constrain the stellar mass (e.g., the determined stellar mass is differed by a factor of 15 with age ranging from $10^{6.45}$ to $10^{6.50}$ years). 

\begin{deluxetable*}{ccc}
\tablenum{3}
\tablecaption{Fixed parameters of the composite SFH model adopted in BEAGLE to generate the mock galaxy spectra.}
\tablewidth{0pt}
\tablehead{
\colhead{Parameter} & \colhead{Value} & \colhead{Description} 
}
\startdata
$\mathrm{log\left(Z_{\ast} / Z_{\odot}\right)}$ & -1.24 & Stellar Metallicity of the composite SFH model  \\
$\mathrm{log\left(Z / Z_{\odot}\right)}$ & -1.24 & Gas-Phase Metallicity of the composite SFH model \\
$\mathrm{log\left(M_{\ast} / M_{\odot}\right)}$ & 6.80  & Total Stellar Mass of the composite SFH model \\
$\mathrm{log(SFR_c/M_{\odot} \ yr^{-1})}$ & -0.85 & SFR of the recent burst in the composite SFH model \\
$\mathrm{\Delta t_{SFR_c} / yr}$ & $10^6$ & Timescale of the recent burst in the composite SFH model  \\
$\mathrm{log \left( U\right)}$ (nebular) & -1.75 & Effective Galaxy-Wide Ionization Parameter \\
$\mathrm{\xi_{d}}$ & 0.1 & Effective Galaxy-Wide Dust-to-Metal Mass Ratio \\
$\mathrm{\mu}$ & 0.3 & Fraction of Attenuation Optical Depth Arising From the Diffuse ISM
\enddata
\tablecomments{Column 1 lists the fixed parameters of BEAGLE in our composite SFH model. Column 2 lists the fixed value of each parameter. The stellar metallicity (row 1) is assumed to be the same as the gas-phase metallicity (row 2) in the ISM. Their values are based on the measured starburst metallicity in this paper. For reference, BEAGLE adopts a solar metallicity $\mathrm{Z_{\odot}} = 0.0152$. The values of stellar mass (row 3) and SFR (row 4) are extracted from the MPA-JHU database. The value of ionization parameter (row 6) is taken from Table 5 in \cite{Olivier_2022}. The value of $\xi_{d}$ is set to be 0.1, the typical value derived in \cite{chevallard_physical_2018} when fitting the combination of \textit{HST}/COS photometry and Balmer emission lines. The value of $\mu$ is the default value from BEAGLE.  Column 3 shows the description of each parameter.}
\label{table:beagle}
\end{deluxetable*}




\vspace{-2.0em}

\subsection{Global Gas Kinematics} \label{subsubsec:velocity_gradient}

We used the strong emission lines ([O\,{\footnotesize II}] $\lambda \lambda 3726, 29$, $\mathrm{H\beta}$, $\mathrm{H\gamma}$, $\mathrm{H\delta}$, and [O\,{\footnotesize III}] $\lambda\lambda 4959,5007$) to measure the spatially extended velocity field. In saturated regions (Section \ref{subsec:data_cube}), all lines except \text{[O\,{\footnotesize III}]} $\lambda 5007$ were fitted, while only \text{[O\,{\footnotesize III}]} $\lambda 5007$ was fitted in unsaturated regions. We first generated a 2D narrow-band image around the strongest emission line, [O\,{\footnotesize III}] $\lambda 5007$, with a target SNR of 10, and then adaptively binned the image in order to define apertures for spectral extraction. Bins failing to meet our threshold SNR of 3 for peculiar velocity were masked.
%

\begin{figure*}[htb]

    \begin{minipage}[b]{0.5\textwidth}
        \includegraphics[width=\textwidth]{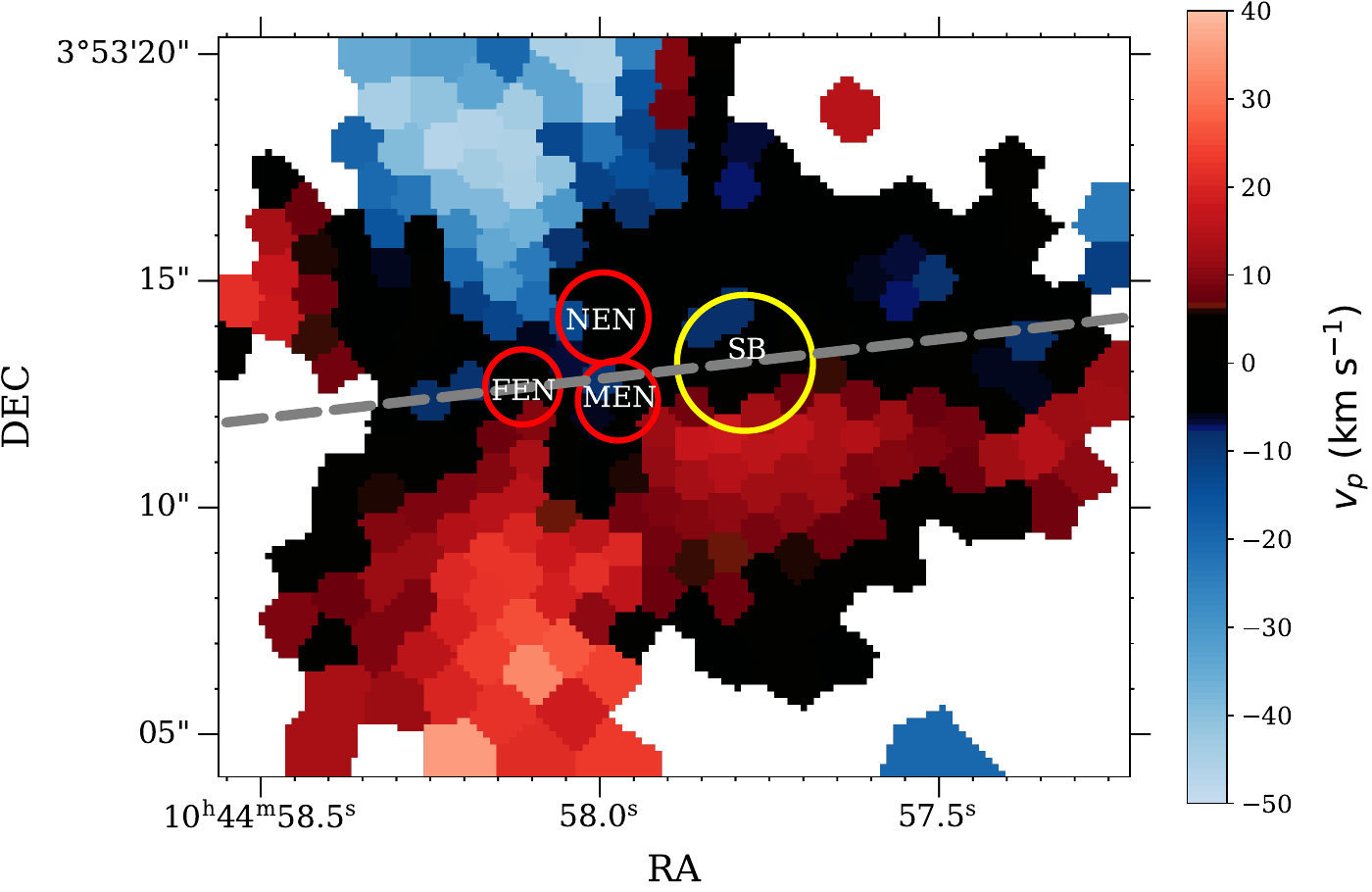}
    \end{minipage}%
    \begin{minipage}[b]{0.5\textwidth}
        \includegraphics[width=\textwidth]{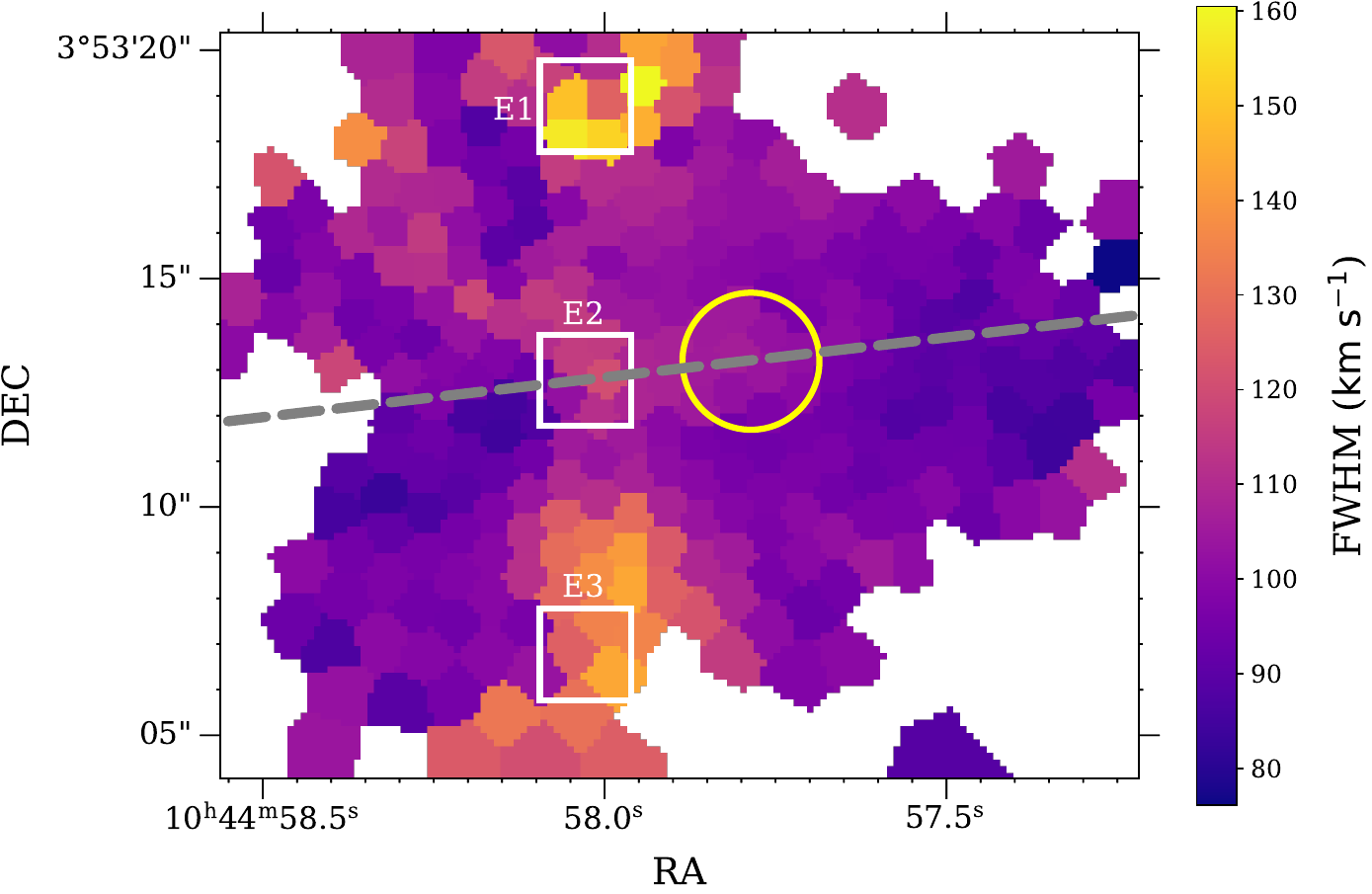}
    \end{minipage}
    \begin{minipage}[b]{\textwidth}
        \centering
        \includegraphics[width=0.5\textwidth]{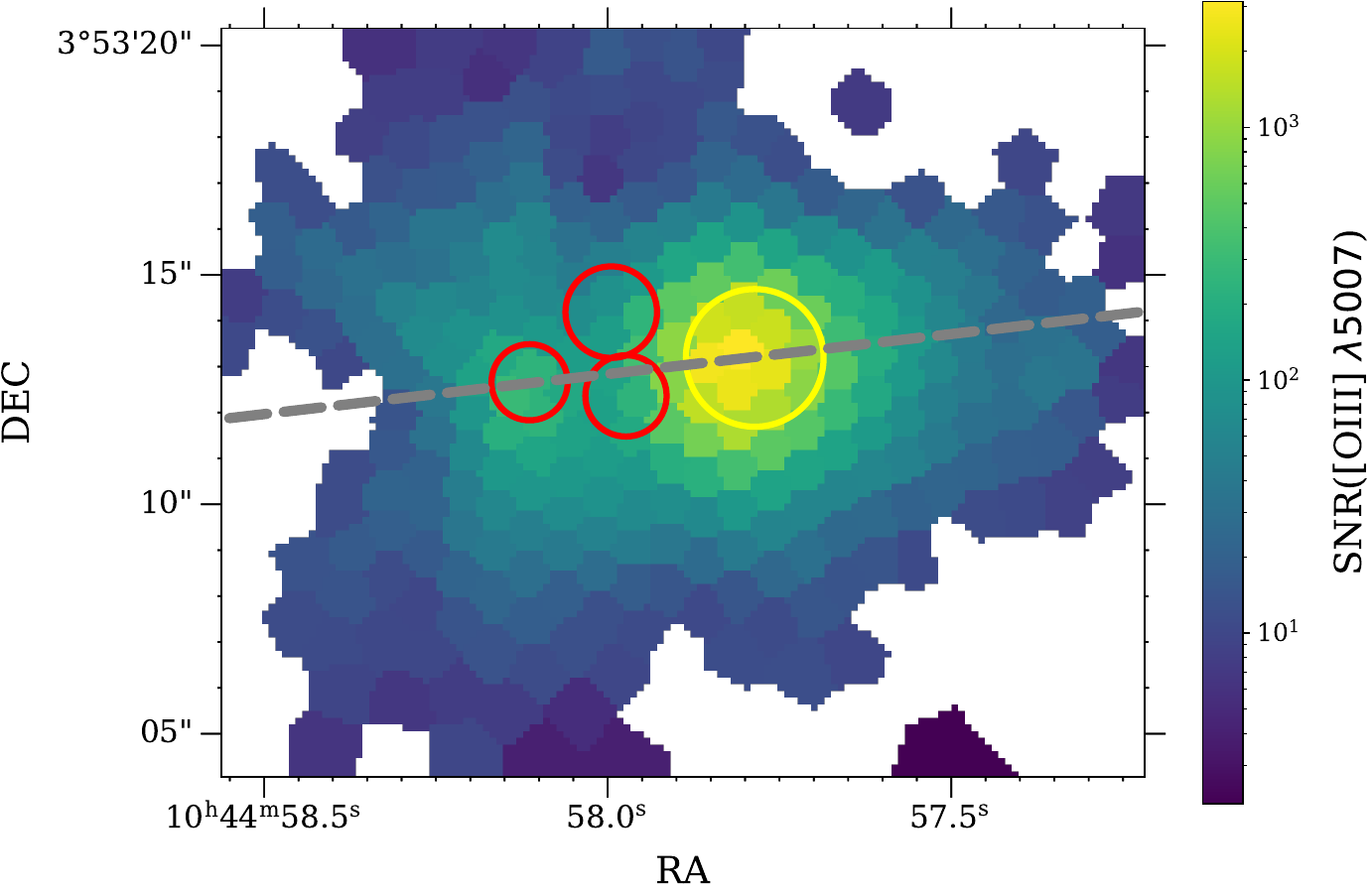}
    \end{minipage}
    \caption{The peculiar velocity map ({\em Top Left}), the FWHM map ({\em Top Right}), and the SNR([O\,{\footnotesize III}]$\lambda 5007$) map  ({\em Bottom Middle}). The bins are obtained via our adaptive binning algorithm with a target SNR of 10 for [O\,{\footnotesize III}]$\lambda 5007$ line. For saturated regions, we simultaneously fit the strong emission lines, [O\,{\footnotesize II}] $\lambda \lambda 3726, 29$, $\mathrm{H\beta}$, $\mathrm{H\gamma}$, $\mathrm{H\delta}$, and [O\,{\footnotesize III}] $\lambda 4959$, to find the peculiar velocity or FWHM. Instead, we only fit the [O\,{\footnotesize III}] $\lambda 5007$ line in the unsaturated regions. {\em Top Left:} Bins with $\text{SNR}(v_p) <$ 3 have been masked out. Each bin is color-coded by its own $v_p$. The bins with reddish color mean they have values of $v_p$ larger than $5 \mathrm{\ km \ s^{-1}}$. The bins with bluish color mean they have values of $v_p$ less than $-5 \mathrm{\ km \ s^{-1}}$. The bins with black color represent they have values of $v_p$ ranging from $-5 \ \mathrm{km \ s^{-1}}$ to $5 \ \mathrm{km \ s^{-1}}$. The position of each stellar association is marked as the way in Figure \ref{fig:j1044_rx10}. 
    {\em Top Right:} Following the same binning procedure, each bin has $\text{SNR}(v_p) \geq 3$. Each bin is color-coded by its own FWHM value in velocity space. 
    The white-square regions with labels `E1', `E2', and `E3' represent the blue-shifted, zero-velocity, and redshifted regions, respectively, in Figure  \ref{fig:asymmetric_line_profiles}. {\em Bottom Middle:} Each bin is color-coded by its SNR of the [O\,{\footnotesize III}]$\lambda 5007$ line. For each panel, the photometric major axis is shown as a grey dashed line. To determine the major axis, the background-subtracted r-band DECalS image (see Section \ref{subsec:source_detection}) is fitted using a 2D Gaussian function that aligns with the second-order moments of the source before the deblending process, implemented with {\tt\string Photutils}. The estimated inclination of this source from {\tt\string Photutils} is $i = \mathrm{cos^{-1}}(b/a) \approx 60^{\circ}$.  \label{fig:velocity_map}}
\end{figure*}

To measure the peculiar velocity, $v_p$, for each bin, we need to measure the peculiar redshift, $z_p$, because $v_p = c z_p$ in the non-relativistic regime. If we define the parameters $z_{\mathrm{single}}$, $z_{\mathrm{all}}$ as the redshifts derived from a single bin and from all the bins respectively, the peculiar velocity of each bin can be determined by the ratio of these two redshifts $v_p = c \left[\frac{1 + z_{\mathrm{single}}}{1 + z_{\mathrm{all}}}- 1\right]$. Following the procedure described above, we obtain the peculiar velocity map shown in the left panel of Figure \ref{fig:velocity_map}. It is clear that most bins along the photometric major axis have $v_p$ ranging from $-5 \mathrm{\ km \ s^{-1}}$ to $5 \mathrm{\ km \ s^{-1}}$ (i.e., black color). We discover a strong velocity gradient parallel to the galaxy's minor axis, with the steepest gradients ($\sim 30 \ \mathrm{km \ s^{-1} \ kpc^{-1}}$; -40 to 50 $\mathrm{km \ s^{-1}}$ over $\pm 1.5$ kpc) appearing to emanate from the oldest stellar association of the post-starburst population (i.e., FEN).

If rotation produces this velocity gradient, then the galaxy is roughly a prolate spheroid rotation about the major axis.  
The possibility of a prolate rotator galaxy is relatively small. As pointed out in \cite{tsatsi_califa_2017}, the prolate rotators are more commonly found among the most massive early-type galaxies (ETGs) ($M_{\ast} \gtrsim 2 \times 10^{11} M_{\odot}$). They claim two progenitors with different disk orientations are responsible for the prolate rotation in the ETGs, a scenario that is unlikely to happen for our low-mass galaxy.

\begin{figure*}[htbp]
\centering
\gridline{\fig{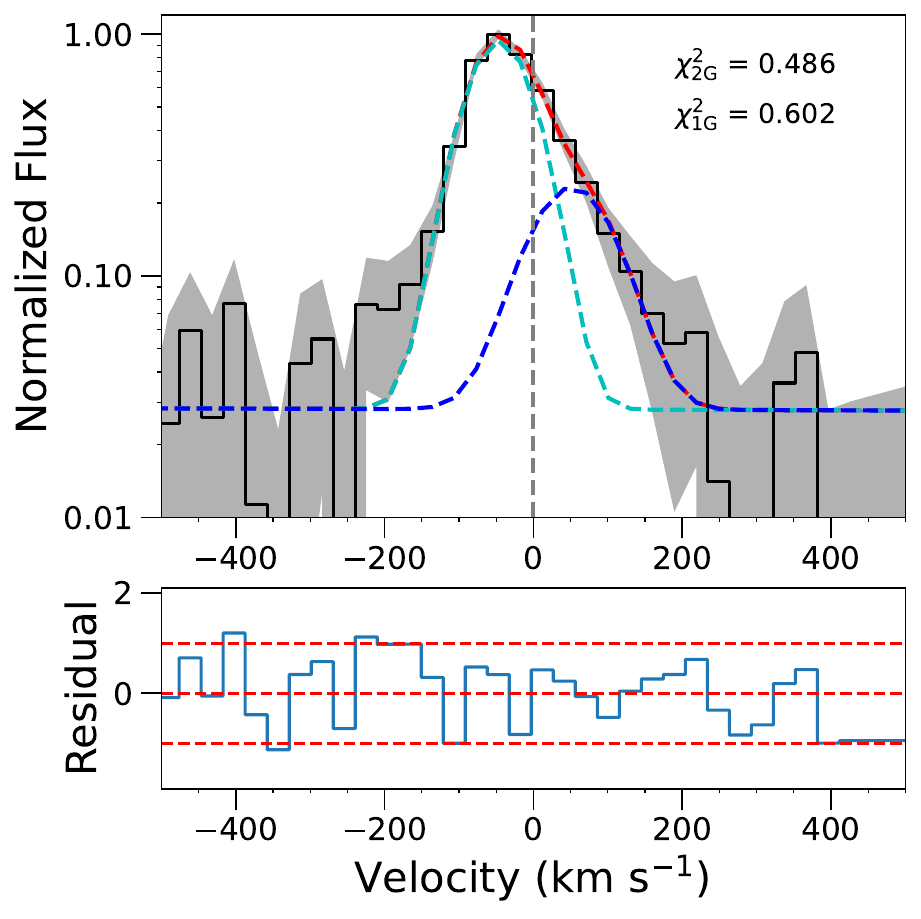}{0.33\textwidth}{\textbf{{\large (E1)}}}
          \fig{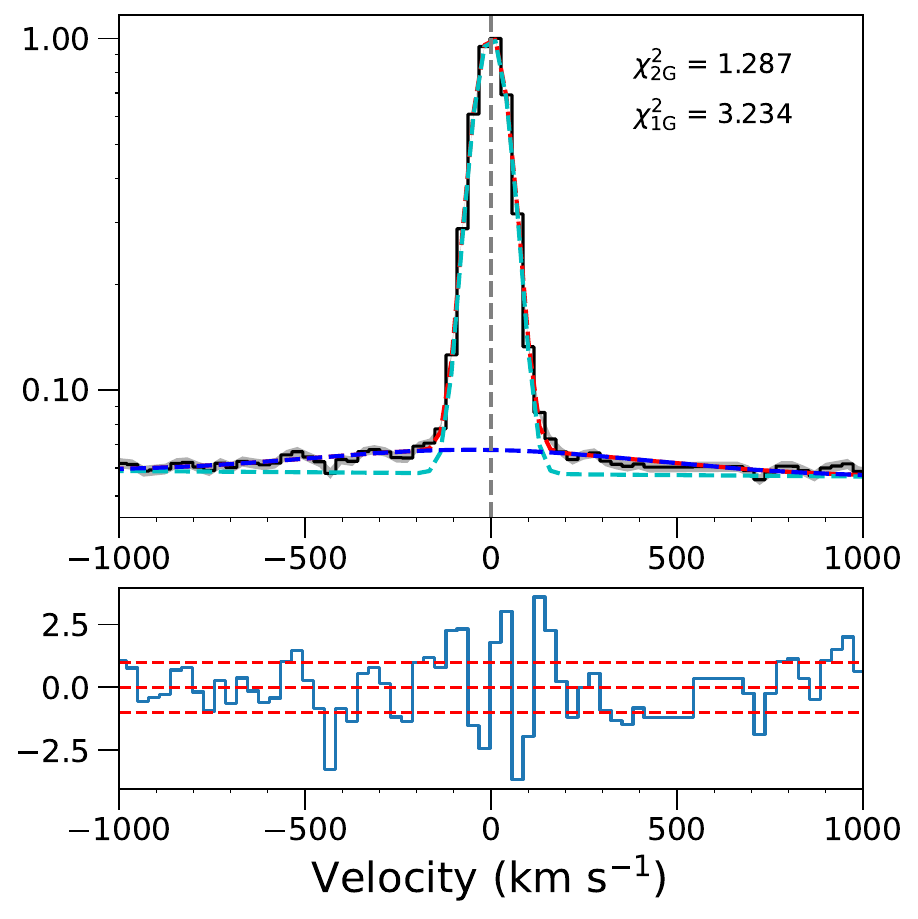}{0.33\textwidth}{\textbf{{\large (E2)}}}
          \fig{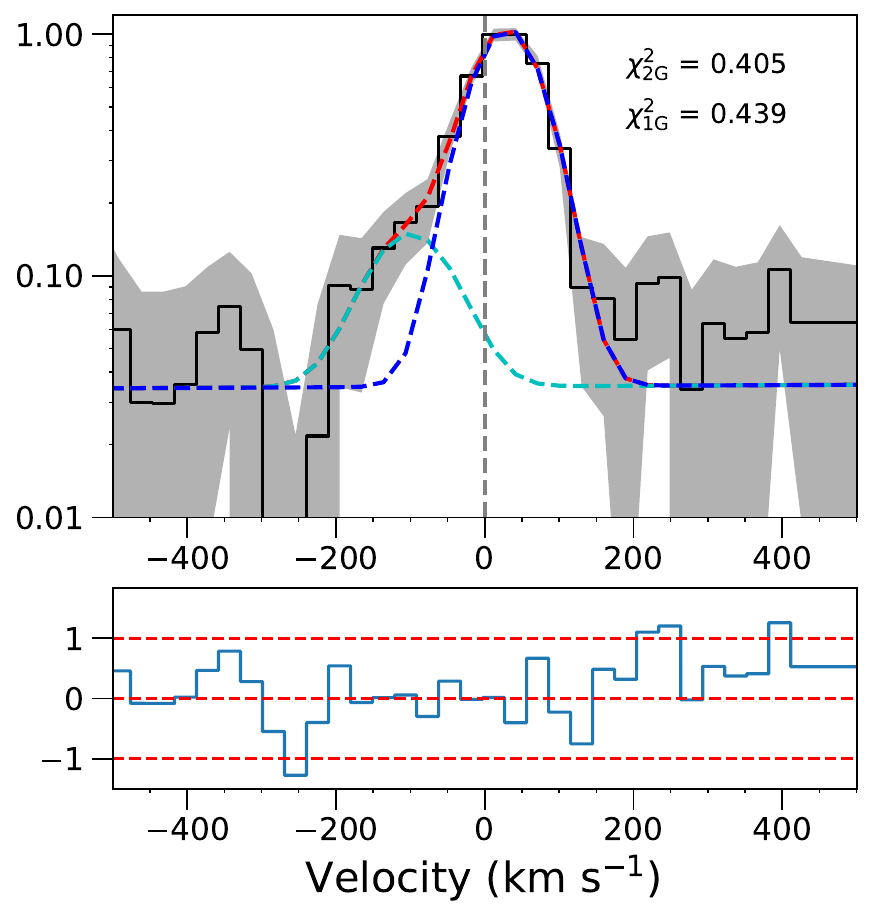}{0.315\textwidth}{\textbf{{\large (E3)}}}
          }
\caption{Variations in [O\,{\footnotesize III}] $\lambda 5007$ line profiles along the minor axis around the post-starburst region. The [O\,{\footnotesize III}] $\lambda 5007$ line is extracted from the E1 or blueshifted ({\em left}), E2 or zero-velocity ({\em middle}), and E3 or redshifted ({\em right}) regions, respectively, in Figure \ref{fig:velocity_map}. Only the double-Gaussian fitting is shown in each panel. In the panels for E1 and E3, the redshift and blueshift components are respectively represented by blue and cyan dashed lines. For E2, the broad and narrow components correspond to the dashed and cyan dashed lines, respectively. Their summation is shown as a red dashed line. The vertical grey dashed line represents the systemic velocity. The reduced $\chi^2$ values for single- and double-Gaussian fittings are listed at the upper right corner of each panel. High absolute peculiar velocities in regions E1 and E3 are associated with an unresolved second velocity component confirmed by the F-test (Section \ref{subsec:line-fitting}). These two velocity component result in asymmetric line profiles and large FWHM values. In the context of a biconical morphology, each component is generated by the front and back sides of the outflow, respectively. The F-test results for E2 suggest a requirement for a second (broad) velocity component. The broad component, indicative of the turbulent outflowing gas, originates from the post-starburst (or the outflow launching) region. Because this broad component shares a velocity centroid similar to the narrow component, it does not significantly impact the line profile to increase its FWHM.
\label{fig:asymmetric_line_profiles}}
\end{figure*}

In contrast, the minor-axis gradient can indicate the existence of galactic outflows \citep[e.g.,][]{Ho_2014}. Based on our age estimations in Section \ref{subsec:ifu_structure}, the post-starburst stellar associations are old enough \citep[i.e., $\gtrsim$ 6 Myr;][]{starburst99} to explode supernovae and provide a specific amount of mechanical energy to power galactic outflows (Martin et al., in prep). In Figure \ref{fig:velocity_map}, regions with high absolute peculiar velocities also have large FWHM values because these broadest lines have asymmetric line profiles (see E1 and E3 regions in Figure \ref{fig:asymmetric_line_profiles}). Higher-resolution spectroscopy would resolve these into two components. The front and back sides of the outflow would each produce a component with different velocity centroids if it follows a biconical morphology \citep{heckman_nature_1990, Lehnert_1996}. Hence the most significant widths represent two distinct components along the sightline rather than the local velocity dispersion.

Besides the existence of galactic outflows, the positive velocity gradient (i.e., the magnitude of $v_p$ increases from the inner regions to the outskirts) tells us it is an accelerating wind instead of a decelerating one. 
Estimating the escape velocity using the relation $v_{\mathrm{esc}} \approx 3 v_{\mathrm{circ}}$ \citep{Veilleux_2020}, and considering $v_{\mathrm{circ}} = 15.6^{+3.0}_{-3.7} \ \mathrm{km \ s^{-1}}$ \citep{xu_classy_2022}, we find that the deprojected velocity $v \approx v_p / \mathrm{cos}(i) \gtrsim 80  \ \mathrm{km \ s^{-1}} $ ($i \approx 60^{\circ}$; see the caption in Figure \ref{fig:velocity_map}) of the largest-velocity bin in either the northern or southern cone exceeds $v_{\mathrm{esc}}$.
Therefore, the outflow may escape from the gravitational potential rather than slowing down and falling back onto J1044+0353.

It is interesting to note the asymmetric velocity field around the starburst region. The top left panel of Figure \ref{fig:velocity_map} shows most of the bins in the north direction of the starburst region have $|v_p|$ less than $5 \ \mathrm{km \ s^{-1}}$. The bins in the south direction, however, have $|v_p|$ ranging from $5 \ \mathrm{km \ s^{-1}}$ to $20 \ \mathrm{km \ s^{-1}}$. These southern bins have relatively symmetric and narrow [O\,{\footnotesize III}] $\lambda 5007$ line profiles, so a single-component fitting is reasonably good. This finding indicates the offset in systematic velocities between these southern bins and the whole galaxy. This offset may reveal the presence of induced inflows that fuel the starburst region during mergers or interactions (Section \ref{subsubsec:potential_explanations}), a hypothesis requiring validation through future H\,{\footnotesize I} mapping. 



\setlength{\tabcolsep}{4.5pt}
\begin{deluxetable}{ccccc}
\tablenum{4}
\tablecaption{Stellar population age and stellar mass of each stellar association.}
\tabletypesize{\footnotesize}
\tablehead{
\colhead{} & STARBURST & MEN & NEN & FEN}
\startdata
log(Age) (year) & $6.480 \pm 0.005$  & $7.2 \pm 0.1$ & $7.2 \pm 0.1$ & $7.30 \pm 0.05$ \\
log($M_\ast$) ($M_\odot$) & $5.82^{+ 0.39}_{- 0.12}$  & $6.26^{+ 0.05}_{- 0.18}$ & $6.08^{+ 0.05}_{- 0.18}$ & $5.57^{+ 0.10}_{- 0.04}$ \\
\enddata
\tablecomments{Row 1 lists the estimated stellar population age for each stellar association on the logarithmic scale. Row 2 lists the estimated stellar mass of each stellar association on the logarithmic scale.}
\label{table:nucleus_age_mass}
\end{deluxetable}


\subsection{Gas-Phase Metallicity} \label{subsec:metallicity}
The temperature-sensitive auroral line [O\,{\footnotesize III}] $\lambda 4363$ is observable in both the starburst and the post-starburst regions, so we can use the ``direct'' method to determine their gas phase oxygen abundances \citep{P_rez_Montero_2017,kewley_2019}. \defcitealias{berg_characterizing_2021}{B21}

Due to the wavelength coverage in KCWI, we can use the line ratio of [O\,{\footnotesize II}] $\lambda \lambda 3726, 29$ doublet to determine electron density. The lower-density limit is set to be 10 $\mathrm{cm^{-3}}$. 
For temperature diagnostics, we utilize the line ratio between [O\,{\footnotesize III}] $\lambda 4363$ line and [O\,{\footnotesize III}] $\lambda\lambda 4959, 5007$ doublet to derive the temperature $\mathrm{T_e}$[O\,{\footnotesize III}]. 
Due to saturation of the [O\,{\footnotesize III}] $\lambda 5007$ line in both long and short exposure cubes of the starburst region (Section \ref{subsec:data_cube}), we estimate this line flux for all regions in J1044+0353 using the intrinsic ratio of the [O\,{\footnotesize III}] $\lambda\lambda 4959,5007$ doublet \citep[$\mathrm{I_{\lambda 5007} / I_{\lambda 4959}}$ = 2.98;][]{SZ_2000}. This method ensures consistency and a physical $\mathrm{T_e}$[O\,{\footnotesize III}] across all regions.

$\mathrm{T_e}$[O\,{\footnotesize III}] is employed to determine the high-ionization zone temperature when measuring the $\mathrm{O^{2+}}$/$\mathrm{H^{+}}$ abundance. Adopting a two-ionization-zone model, we also need to know the ionic abundance of $\mathrm{O^{+}}$/$\mathrm{H^{+}}$ (i.e., $\mathrm{O}$/$\mathrm{H}$ = $\mathrm{\left(O^{+} + O^{2+}\right)}$/$\mathrm{H^{+}}$). This assumption is valid because  \citetalias{berg_characterizing_2021} apply a four-ionization-zone model to determine the metallicity of J1044+0353 and find the fraction of oxygen in the neutral and very-high ionization states is insignificant (i.e., $\mathrm{(O^{0} + O^{3+}) / O} \sim$ $4 \%$). Therefore, it is reasonable for us to use the simplified two-zone model here. Because most temperature-sensitive [O\,{\footnotesize II}] emission lines lie in the UV region or in the red-end of the optical region (e.g., [O\,{\footnotesize II}] $\lambda \lambda 7320, 7330$), a scaling relation between $\mathrm{T_e}$[O\,{\footnotesize III}] and $\mathrm{T_e}$[O\,{\footnotesize II}] is needed to determine the ionic abundance of $\mathrm{O^{+}}$/$\mathrm{H^{+}}$. 


We apply temperature relations proposed in \cite{arellano-cordova_ten_2020} \defcitealias{arellano-cordova_ten_2020}{AR20} (hereafter AR20) to determine metallicity for each stellar association or each bin (in Section \ref{subsubsec:oh_gradient}).
We choose not to use the relation of \cite{garnett_electron_1992} (see also \cite{campbell_stellar_1986}) adopted in \citetalias{berg_carbon_2016} because this relation tends to overestimate the temperature of the low-ionization zone for galaxies with low metallicities \citepalias[$\mathrm{12 + log(O/H) \lesssim 7.6}$; Figure 3 in][]{arellano-cordova_ten_2020}.
Owing to this fact, \cite{garnett_electron_1992} might underestimate the ionic abundance of $\mathrm{O^{+}}$ in the low-metallicity environment like J1044+0353 that it does not depend on the $P$ = $\mathrm{I_{[O\,{\footnotesize III}] \lambda \lambda 4959, 5007}}$ / ($\mathrm{I_{[O\,{\footnotesize III}] \lambda \lambda 4959, 5007}}$ + $\mathrm{I_{[O\,{\footnotesize II}] \lambda 3726}}$) parameter defined in \citetalias{arellano-cordova_ten_2020}.

Given the temperatures of these two ionization zones, we can derive the ionic abundance of $\mathrm{O^{+}/H^{+}}$ or $\mathrm{O^{2+}/H^{+}}$ using its calibration with the relative intensity of [O\,{\footnotesize II}] $\lambda\lambda 3726, 3729$ or [O\,{\footnotesize III}] $\lambda\lambda 4959, 5007$ doublet to H$\beta$. 
Following a similar approach as \citet{P_rez_Montero_2017}, these calibrations are obtained from fittings to {\tt\string PyNeb} v1.1.14 using the default collision strengths of $\mathrm{O^{+}}$ and $\mathrm{O^{2+}}$ \citep{kisielius_electron-impact_2009,Storey_2014}. These two relations are given in Equations \ref{3.1} and \ref{3.2} with $t_l = 10^{-4} \  \mathrm{T_e[O\,{\footnotesize II}]}$ and $t_h = 10^{-4} \ \mathrm{T_e[O\,{\footnotesize III}]}$. These two new calibrations have precisions better than 0.01 dex in the temperature range $\mathrm{T_e}$[O\,{\footnotesize III}] = $\mathrm{7,000 \ K - 25,000 \ K}$ and in the density range $\mathrm{n_e} = \mathrm{10 \ cm^{-3} - 1000 \ cm^{-3}}$.
\begin{eqnarray}
\label{3.1} 12 + \mathrm{log \left(\frac{O^+}{H^+}\right)} = \mathrm{log \left(\frac{I_{\mathrm{[O\,{\footnotesize II}] \ \lambda3726}} + I_{\mathrm{[O\,{\footnotesize II}] \ \lambda3729}}}{I_{\mathrm{H\beta}}}\right)} 
\nonumber \\ + 5.90  + \frac{1.65}{t_l} 
    -0.54 \cdot \mathrm{log} \left(t_l \right) + 7.99 \times 10^{-5} \cdot n_e \  
\end{eqnarray}

\begin{eqnarray}
    \label{3.2} 12 + \mathrm{log \left(\frac{O^{2+}}{H^{+}}\right)} = \mathrm{log \left(\frac{I_{\mathrm{[O\,{\footnotesize III}] \ \lambda4959}} + I_{\mathrm{[O\,{\footnotesize III}] \ \lambda5007}}}{I_{\mathrm{H\beta}}}\right)} \nonumber \\  + 6.21 + 
    \frac{1.21}{t_h}-0.656 \cdot \mathrm{log} \left(t_h \right) \ \ \ \ 
\end{eqnarray}
\setlength{\tabcolsep}{12pt} 
\vspace{-0.5cm} 
\begin{deluxetable*}{ccccc}
\tablenum{5}
\tablecaption{Physical Properties of Different Stellar Associations}
\tablehead{
\colhead{} & \colhead{SB} & \colhead{MEN} & \colhead{NEN} & \colhead{FEN} 
}
\startdata
\colhead{(1) $\mathrm{I_{\mathrm{[O\,{\footnotesize II}] \ \lambda3726}}}$} & 71.41$^{+ 1.96}_{- 1.96}$ & 1.54$^{+ 0.15}_{- 0.15}$ & 0.47$^{+ 0.05}_{- 0.05}$ & 2.67$^{+ 0.05}_{- 0.05}$  \\
\colhead{(2) $\mathrm{I_{\mathrm{[O\,{\footnotesize II}] \ \lambda3729}}}$} & 91.18$^{+ 2.10}_{- 2.10}$ & 2.08$^{+ 0.15}_{- 0.15}$ & 0.67$^{+ 0.05}_{- 0.05}$ & 4.03$^{+ 0.05}_{- 0.05}$ \\
(3) $\mathrm{n_e}$[O\,{\footnotesize II}] ($\mathrm{cm^{-3}}$) & 193$^{+ 34}_{- 33}$ & 119$^{+ 110}_{- 76}$ & 45$^{+ 92}_{- 45}$ & 10$^{+ 10}_{- 10}$ \\
\colhead{(4) $\mathrm{I_{\mathrm{[O\,{\footnotesize III}] \ \lambda4363}}}$ (narrow)}  & 68.84$^{+1.39}_{-1.39}$ & 0.66$^{+ 0.04}_{- 0.04}$ & 0.21$^{+ 0.02}_{- 0.02}$ & 0.42$^{+ 0.02}_{- 0.02}$ \\
\colhead{(5) $\mathrm{I_{\mathrm{[O\,{\footnotesize III}] \ \lambda4363}}}$ (broad)} & 5.70$^{+3.66}_{-3.66}$ & \nodata & \nodata & \nodata \\
\colhead{(6) $\mathrm{I_{\mathrm{[O\,{\footnotesize III}] \ \lambda4959}}}$ (narrow)} & 727.67$^{+ 4.12}_{- 4.12}$ & 5.98$^{+ 0.06}_{- 0.06}$ & 2.40$^{+ 0.03}_{- 0.03}$ & 5.36$^{+ 0.03}_{- 0.03}$ \\
\colhead{(7) $\mathrm{I_{\mathrm{[O\,{\footnotesize III}] \ \lambda4959}}}$ (broad)} & 31.43$^{+18.86}_{-18.86}$ & \nodata & \nodata & \nodata \\
(8) $\mathrm{T_e}$[O\,{\footnotesize III}] (K) & 19313$^{+ 158}_{- 157}$ & 21241$^{+ 644}_{- 569}$ & 18420$^{+ 632}_{- 583}$ & 17272$^{+ 348}_{- 326}$  \\
(9) $\mathrm{12 + log(O/H)}$ & $7.45 \pm 0.01$ & $7.37 \pm 0.03$ & $7.52 \pm 0.03$ & $7.43 \pm 0.03$  \\
\colhead{(10) $\mathrm{I_{H\beta}}$ (narrow)} & 502.28$^{+ 3.22}_{- 3.22}$ & 4.91$^{+ 0.05}_{- 0.05}$ & 1.71$^{+ 0.03}_{- 0.03}$ & 6.95$^{+ 0.04}_{- 0.04}$ \\
\colhead{(11) $\mathrm{I_{H\beta}}$ (broad)}  & 21.49$^{+ 13.02}_{- 13.02}$ & \nodata & \nodata & \nodata \\
\colhead{(12) $\mathrm{I_{H\gamma}}$} & 253.02$^{+ 2.27}_{- 2.27}$ & 2.33$^{+ 0.05}_{- 0.05}$ & 0.81$^{+ 0.03}_{- 0.03}$ & 3.31$^{+ 0.03}_{- 0.03}$ \\
\colhead{(13) $\mathrm{I_{H\delta}}$} & 141.76$^{+ 1.89}_{- 1.89}$ & 1.16$^{+ 0.06}_{- 0.06}$ & 0.35$^{+ 0.03}_{- 0.03}$ & 1.81$^{+ 0.03}_{- 0.03}$ \\
(14) $\mathrm{EW \ (EM)_{\mathrm{H\beta}}}$ & $299 \pm 30$ & $6.82 \pm 0.69$ & $5.51 \pm 0.56$ & $46.02 \pm 4.67$  \\
(15) $\mathrm{EW \ (AB)}_{\mathrm{H\beta}}$ & \nodata & $2.94 \pm 0.20$ & $2.68 \pm 0.18$ & $2.58 \pm 0.15$ \\
(16) $\mathrm{EW \ (EM)}_{\mathrm{H\gamma}}$ & $105 \pm 10$ & $2.22 \pm 0.21$ & $1.86 \pm 0.18$ & $16.59 \pm 1.58$ \\
(17) $\mathrm{EW \ (AB)}_{\mathrm{H\gamma}}$ & \nodata & $4.41 \pm 0.28$ & $4.89 \pm 0.31$ & $4.14 \pm 0.23$ \\
(18) $\mathrm{EW \ (EM)}_{\mathrm{H\delta}}$ & $49 \pm 4$ & $0.91 \pm 0.08$ & $0.67 \pm 0.06$ & $7.56 \pm 0.65$ \\
(19) $\mathrm{EW \ (AB)}_{\mathrm{H\delta}}$ & \nodata & $4.50 \pm 0.27$ & $4.02 \pm 0.24$ & $4.94 \pm 0.27$  \\
(20) $A_V$ & 0.20$^{+ 0.10}_{- 0.10}$ & 0.56$^{+ 0.09}_{- 0.09}$ & 0.21$^{+ 0.13}_{- 0.12}$ & 0.20$^{+ 0.04}_{- 0.04}$ 
\enddata
\tablecomments{The first two rows show the intensities of [O\,{\footnotesize II}] $\lambda\lambda 3726, 3729$ doublet, whose ratio is used to derive the electron density $\mathrm{n_e}$ of each stellar association in row 3. The lower-density limit is set to be 10 $\mathrm{cm^{-3}}$. 4th and 5th (6th and 7th) rows show the intensities of narrow and broad components of [O\,{\footnotesize III}] $\lambda 4363$ ([O\,{\footnotesize III}] $\lambda 4959$) line. The narrow components of these two [O\,{\footnotesize III}] lines are used to determine the electron temperature $\mathrm{T_e}$ of each stellar association in row 8. The intensity of [O\,{\footnotesize III}] $\lambda 5007$ line is not shown here because its intensity is based on that of [O\,{\footnotesize III}] $\lambda 4959$ line (see Section \ref{subsec:metallicity}). The gas phase oxygen abundance for each stellar association is shown in the 9th row. Rows 10 and 11 show the intensities of narrow and broad components of H$\beta$ line. Rows 12 and 13 list the intensities of $\mathrm{H\gamma}$ and $\mathrm{H\delta}$. The equivalent widths, including both emission (EW (EM)) and absorption (EW (AB)) portions, of Balmer lines are shown from the rows 14 to 19. The intrinsic magnitude of extinction, $A_V$, is derived from the Balmer Decrement in the 20th row. All the intensities shown have been corrected for both Galactic and intrinsic extinction. The intensity units is $10^{-16} \mathrm{\ erg \ cm^{-2} \ s^{-1}}$. \label{table:j1044_physical}}
\end{deluxetable*}

\subsubsection{Starburst Metallicity} \label{subsubsec:sb_metallicity}

Following the direct method described in Section \ref{subsec:metallicity}, we can determine the electron density $\mathrm{n_e}$, electron temperature $\mathrm{T_e}$[O\,{\footnotesize III}], and oxygen abundance for the starburst region. The second column of Table \ref{table:j1044_physical} summarizes these physical properties of the starburst region. 



The $\mathrm{T_e}$[O\,{\footnotesize III}] temperature, however, cannot represent the temperature across the ionization zones. Previous studies use different approaches to tackle this issue. \citetalias{berg_carbon_2016} utilize the calibration from \cite{garnett_electron_1992} to determines the low ionization zone temperature from $\mathrm{T_e}$[O\,{\footnotesize III}], which could underestimate the ionic abundance of $\mathrm{O^{+}}$ (Section \ref{subsec:metallicity}). If the galaxy is highly excited, then the impact should be negligible because the contribution of $\mathrm{O^{+}}$ to the total oxygen abundance is subdominant. Due to this difference, even if we get similar electron density and temperature as \citetalias{berg_carbon_2016}, we find a relatively higher ionic abundance $\mathrm{O^{+}/H^{+}}$ (Table \ref{table:oh_comparison}). \citetalias{berg_characterizing_2021}, instead, use different strong line ratios to determine $\mathrm{T_e}$ in different ionization zones (e.g., the [O\,{\footnotesize II}] $\lambda \lambda 7320,30 / \lambda \lambda 3726,29$ line ratio is used in the low-ionization zone).

Other than the temperature calibration, there were subtle differences between our ``direct'' method and those used in previous work. \citetalias{berg_carbon_2016} utilizes the density-sensitive doublet [O\,{\footnotesize II}] $\lambda \lambda 7320,30$ instead of [O\,{\footnotesize II}] $\lambda \lambda 3726,29$ used in this study. \citetalias{berg_characterizing_2021} use [S\,{\footnotesize II}] $\lambda\lambda 6717, 6731$, Si\,{\footnotesize III}] $\lambda\lambda 1883, 1892$, C\,{\footnotesize III}] $\lambda\lambda 1907, 1909$, and Ar\,{\footnotesize IV}] $\lambda\lambda 4711, 4740$ line ratios to determine $\mathrm{n_e}$ in their four ionization zones respectively. 
We illustrate the differences in determining intrinsic reddening values in Section \ref{subsec:extinction}. 
Moreover, the methods used by \citetalias{berg_carbon_2016} and \citetalias{berg_characterizing_2021} to determine the oxygen abundance rely on SDSS and LBT/MODS spectra, respectively. However, according to \citet{arellano-cordova_classy_2022}, the differences between the long-slit and IFU spectra are within the uncertainties when the high-ionization zone temperature $\mathrm{T_e}$[O\,{\footnotesize III}] is measured directly. These facts lead to the negligible deviation between our result and theirs ($\lesssim 0.01$ dex; Table \ref{table:oh_comparison}). 

\subsubsection{Metallicity in the Post-starburst Region} \label{subsubsec:older_nuclei}
 We found that all eastern stellar associations (NEN: $7.52^{+0.03}_{-0.03}$, MEN: $7.37^{+0.03}_{-0.03}$, and FEN: $7.43^{+0.03}_{-0.03}$) have metallicities within three standard deviations of the starburst metallicity in terms of 12 + log(O/H). 
 Overall, the stellar-mass-weighted metallicity of the post-starburst region ($7.43^{+0.02}_{-0.02}$) is comparable to that of the starburst region. 

The finding that these older stellar associations have similar metallicities as the starburst region is not surprising. Although the post-starburst region is old enough to explode supernovae and expel the produced metals to the ISM, this ejecta might reside in the hot ionized gas (i.e., T $\sim 10^6$ K), which needs to be traced by X-ray emissions  \citep{Martin_2002}.
This gas would remain unmixed with the $\sim 10^4$ K gas until the radiative cooling is significant to detect these metals in the warm phase of the ISM \citep[i.e., T $\sim 10^4$; K][]{tenorio-tagle_interstellar_1996}. 

%
Due to gravity, this gas may eventually fall onto the galactic disk and can be mixed with the ISM matter when the next generation of stars photoionize this gas. Consequently, these processes might take $\sim 100$ Myr for the chemical enrichment by the metal-rich gas to happen \citep{ST07}. This mixing timescale is one order of magnitude larger than the estimated age of the post-starburst region. Therefore, this ejecta is subject to a long mixing timescale to be well-mingled with the matters in the ISM. 

\vspace{-0.5cm} 
\begin{deluxetable*}{ccccccc}
\tablenum{6}
\tablecaption{Comparison to Previous Studies on J1044+0353}
\tablewidth{0pt}
\tablehead{
\colhead{} & \colhead{$\mathrm{n_e}$ ($\mathrm{cm^{-3}}$)} & \colhead{$\mathrm{T_e}$[O\,{\footnotesize II}] (K)} & \colhead{$\mathrm{T_e}$[O\,{\footnotesize III}] (K)} &
\colhead{$\mathrm{O^{+} / H^{+}} \ (10^5)$} & \colhead{$\mathrm{O^{2+} / H^{+}} \ (10^5)$} & \colhead{$\mathrm{12 + log(O/H)}$} 
}
\startdata
STARBURST & 193$^{+ 34}_{- 33}$ & 14796$^{+ 988}_{- 988}$ & 19313$^{+ 158}_{- 157}$ & 0.28$^{+ 0.06}_{- 0.06}$ & 2.56$^{+ 0.05}_{- 0.05}$ & $7.45 \pm 0.01$ \\
\citetalias{berg_carbon_2016} & 260 & 16300$\pm 300$ & 19600$\pm 500$ & 0.17$\pm 0.10$ & 2.62$\pm 0.03$  & $7.45 \pm 0.02$ \\
\citetalias{berg_characterizing_2021} & $200 \pm 40$ & 19100$\pm 1500$ & 19200$\pm 200$ & 0.114$\pm 0.025$ & 2.70$\pm 0.067$ & 7.44$\pm 0.01$ \\
\enddata
\tablecomments{Comparison of $\mathrm{n_e}$, $\mathrm{T_e}$[O\,{\footnotesize II}], $\mathrm{T_e}$[O\,{\footnotesize III}], and gas phase oxygen abundance with previous studies conducted by \citetalias{berg_carbon_2016} and \citetalias{berg_characterizing_2021}. Row 1 shows these physical quantities of the starburst region obtained in this study (Table \ref{table:j1044_physical}). Row 2 is extracted from Table 3 in \citetalias{berg_carbon_2016}. Row 3 is extracted from Table 3 and 5 from \citetalias{berg_characterizing_2021}, which is based on a four-ionization-zone model. The $\mathrm{n_e}$ in our study is based on [O\,{\footnotesize II}] $\lambda\lambda 3726, 3729$ line ratio; however, the $\mathrm{n_e}$ in \citetalias{berg_carbon_2016} is based on [S\,{\footnotesize II}] $\lambda\lambda 6717, 6731$ line ratio. \citetalias{berg_characterizing_2021}, instead, use [S\,{\footnotesize II}] $\lambda\lambda 6717, 6731$, Si\,{\footnotesize III}] $\lambda\lambda 1883, 1892$, C\,{\footnotesize III}] $\lambda\lambda 1907, 1909$, and Ar\,{\footnotesize IV}] $\lambda\lambda 4711, 4740$ line ratios to determine $\mathrm{n_e}$ in different ionization zones. 
The low-ionization-zone temperature $\mathrm{T_e}$[O\,{\footnotesize II}] is estimated using the \citetalias{arellano-cordova_ten_2020}  (or \citet{garnett_electron_1992}) calibration between $\mathrm{T_e}$[O\,{\footnotesize II}] and $\mathrm{T_e}$[O\,{\footnotesize III}] in our study (or in \citetalias{berg_carbon_2016}). \citetalias{berg_characterizing_2021} directly measure $\mathrm{T_e}$[O\,{\footnotesize II}] by using the [O\,{\footnotesize II}] $\lambda \lambda 7320,30 / \lambda \lambda 3726,29$ line ratio.
}
\label{table:oh_comparison}
\end{deluxetable*}


\subsubsection{Chemical Inhomogeneities} \label{subsubsec:oh_gradient}

\begin{figure*}[htb]
    \begin{minipage}[b]{0.55\textwidth}
        \includegraphics[width=\textwidth]{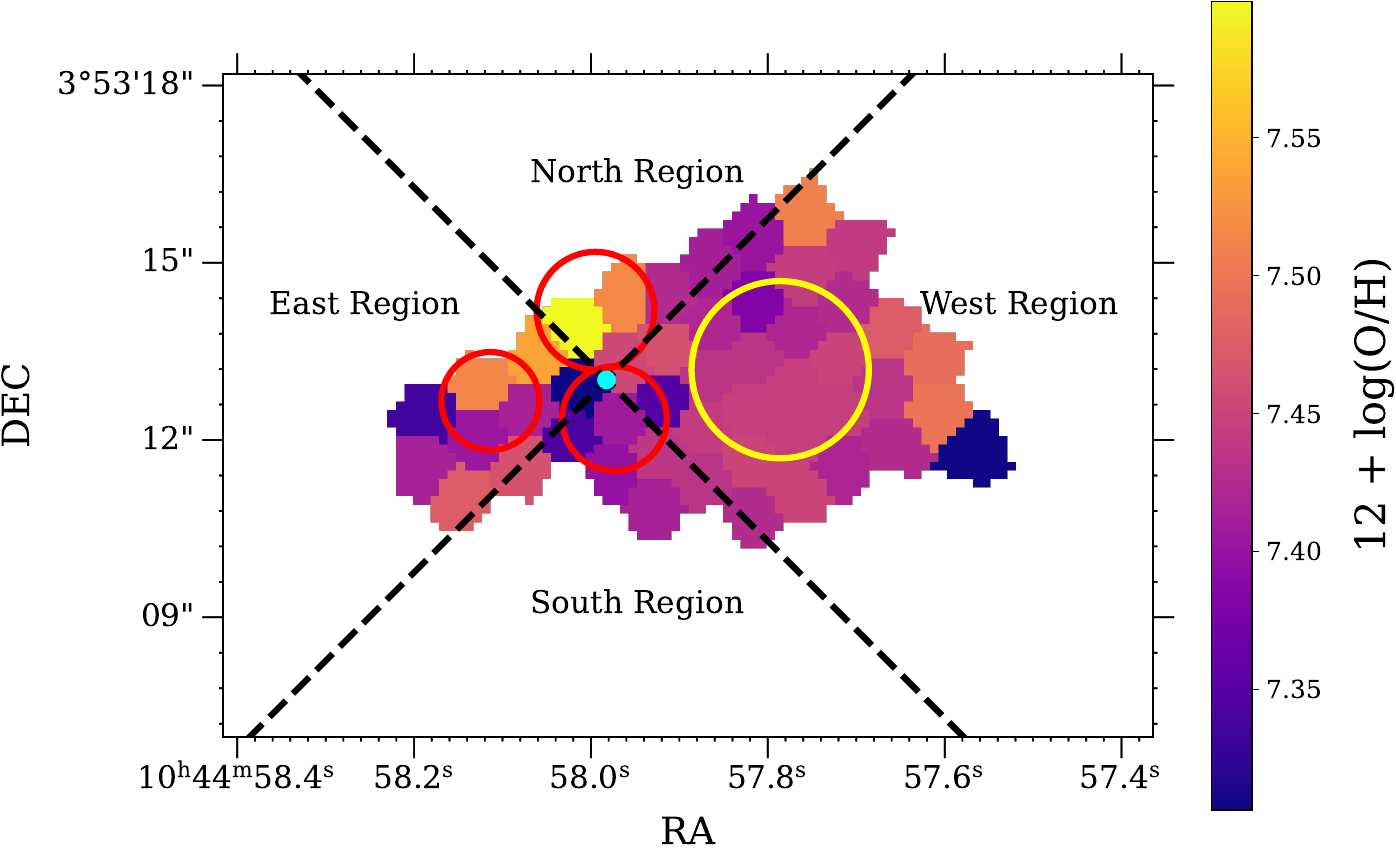}
    \end{minipage}
    \begin{minipage}[b]{\textwidth}
        \includegraphics[width=0.45\textwidth]{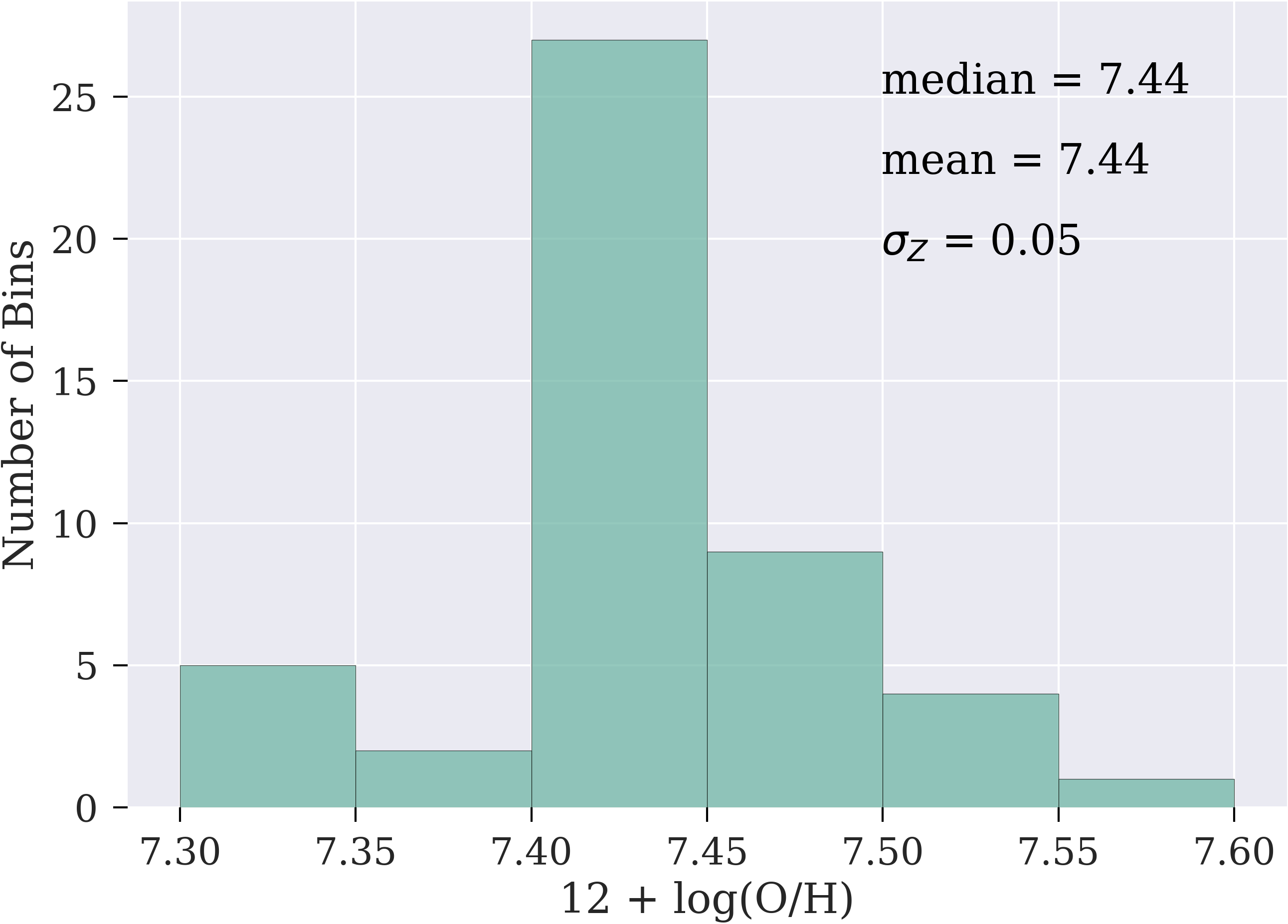}
    \end{minipage}
\caption{
{\em Left:} The gas-phase metallicity map with the metallicities measured from the direct method. The bins are obtained via our adaptive binning algorithm with a target SNR = 5 for [O\,{\footnotesize III}] $\lambda 4363$ line. This map is divided into four regions (by two black dashed lines) with different position angles so we can find out how metallicity gradients change with position angles (shown in Figure \ref{fig:oh_gradient}). The crossing of these two black dashed lines is the stellar-mass-weighted center (marked by the cyan circular point). Each bin is color-coded by its own metallicity in terms of 12 + log(O/H). 
{\em Right:} Distribution of the gas-phase metallicity of each bin obtained by the direct method. The median, mean, and standard deviation of the metallicity distribution are shown in the upper right corner. 45 out of 48 bins have metallicities within two standard deviations ($\sigma_{\mathrm{Z}} = 0.05$) of the median value. Two of the other three bins have metallicities of 7.31, and one has a metallicity of 7.60.}  \label{fig:oh_map}
\end{figure*}

The majority of research indicates chemical homogeneity in ionized gas at scales $\gtrsim 100$ pc with spreads of $\lesssim 0.2$ dex \citep[e.g.,][]{KS_1997,LS_2004,Kumari_2019}. It is crucial, however, to note that this argument does not hold in every instance \citep[e.g.,][]{James_2020, FD_2023}.
Here we examine the chemical inhomogeneities of J1044+0353 by presenting the direct-based metallicity as a function of distance from the stellar-mass-weighted center of the four stellar associations (the cyan circular point in the left panel of Figure \ref{fig:oh_map}).

We applied the adaptive binning to a 2D narrow-band image of the faintest line, [O\,{\footnotesize III}] $\lambda 4363$, required by the oxygen abundance diagnostic at a SNR threshold of 5. We obtained 48 bins in total, shown in Figure \ref{fig:oh_map} (and in Figure \ref{fig:balmer_ratio_map}). After applying the binning algorithm to the 2D image, we follow the procedure outlined in Section \ref{subsec:metallicity} to determine the metallicity of each bin. To address three bins that show unphysical extinction values using the intrinsic Case A ratio (white contours in Figure \ref{fig:balmer_ratio_map}), we assign these bins with $A_V = 0$ in the metallicity measurements.

We first show the metallicity distribution of all bins in the right panel of Figure \ref{fig:oh_map}. To quantify the size of the spread in metallicity, we calculate an interquartile range (IQR) of 0.04. 45 out of 48 bins have metallicities within two standard deviations ($\sigma_{\mathrm{Z}} = 0.05$) of the median value (12 + log(O/H) = 7.44). Among the three outliers, two possess metallicities of 7.31, and one presents a metallicity of 7.60, signifying the largest spread of approximately 0.3 dex between individual bins. Although J1044+0353 is considered chemically homogeneous due to its relatively small IQR and $\sigma_{\mathrm{Z}}$, the existence of spatial variations among these outlier regions, with a resolution element size of approximately 250 pc, cannot be disregarded, particularly in the post-starburst region where two of these outliers are located within this area.

\begin{figure*}
\centering
\includegraphics[width=1.0\textwidth]{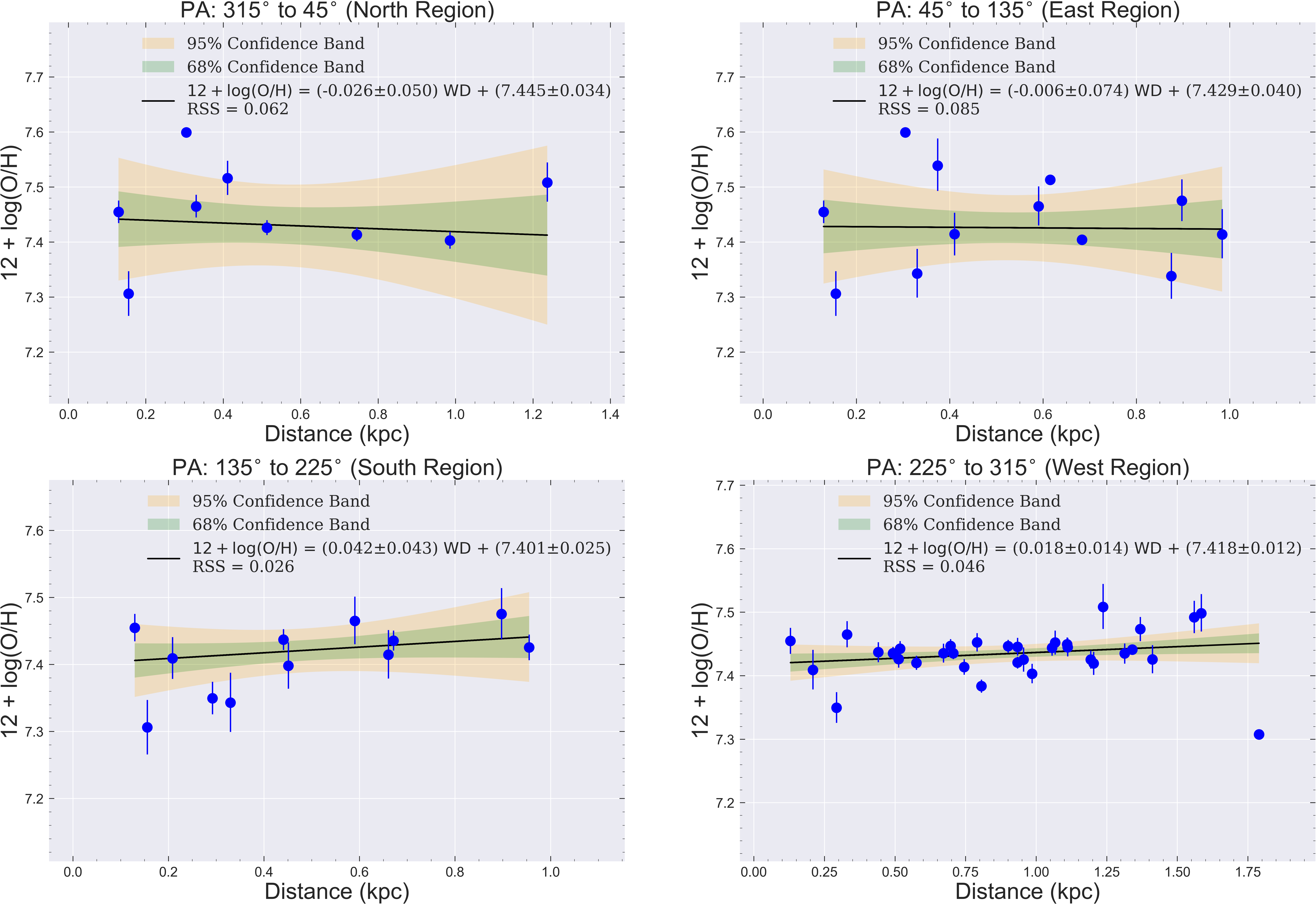}
\caption{Metallicity gradients in different position angles. The galaxy is divided into four regions with corresponding position angles (shown in the title of each panel; see Figure \ref{fig:oh_map}). For each panel, the black solid line is the best-fitting linear function. The RSS (residual sum of squares) of each best-fitting linear model is presented in each panel. The green and orange areas represent the $68 \%$ and the $95 \%$ confidence bands of the best-fitting line. The gradients in the South and West Regions could potentially be positive; however, their error bars are comparable to their respective values. The North and East Regions, centered around the post-starburst region, show slightly larger scatter compared to the other two regions. Overall, all regions exhibit relatively flat metallicity gradients considering error bars, indicating minor azimuthal metallicity variations. } \label{fig:oh_gradient}
\end{figure*}

We then divide the galaxy into four regions to see if the metallicity distributions vary with position angles. We present metallicity as a function of the distance to the spaxel closest to the stellar-mass-weighted center.
The distance of each bin corresponds to the median projected distance of spaxels allocated to that bin.
We find metallicity gradients are ranging from $-0.026 \pm 0.050$ to $0.042 \pm 0.043$ dex $\mathrm{kpc^{-1}}$ across four regions (Figure \ref{fig:oh_gradient}). The gradients in the South and West Regions could potentially be positive; however, their error bars are comparable to their respective values. The North and East Regions, centered around the post-starburst region, show slightly larger scatter compared to the other two regions (i.e., quantified by the values of residual sum of squares, or RSS, in Figure \ref{fig:oh_gradient}), suggesting potential chemical inhomogeneities along these two position angles. Overall, all regions exhibit relatively flat metallicity gradients considering error bars, indicating minor azimuthal metallicity variations.
If we present metallicity as a function of the distance to the center of the starburst region, the offsets in metallicity gradients (along all PAs) are well within the uncertainties.

\section{Discussion} \label{sec:discussion}
Our stellar population age measurements (Section \ref{subsec:ifu_structure}) and the velocity map (Figure \ref{fig:velocity_map}) reveal the existence of a galactic outflow that appears to emanate from the post-starburst region. The metallicity map (Figure \ref{fig:oh_map}) and the corresponding gradients along different PAs (Figure \ref{fig:oh_gradient}), instead, reveal a considerable scatter of metallicities within the post-starburst region, indicating an induced inflow of low-metallicity gas by a merger or an interaction event. To better interpret the role of each mechanism, we can plot the relative positions of J1044+0353 to the mass-metallicity relation (MZR) and the SFR-stellar mass relation (SFMR) (Section \ref{subsec:offset_MZR_SFMR}). We then discuss each mechanism respectively and quantify their impacts in this galaxy (Section  \ref{subsec:spatial_variation_oh}).

\subsection{Offset from MZR and SFMR} \label{subsec:offset_MZR_SFMR}

\begin{figure*}[htb]
\includegraphics[width=1.0\textwidth]{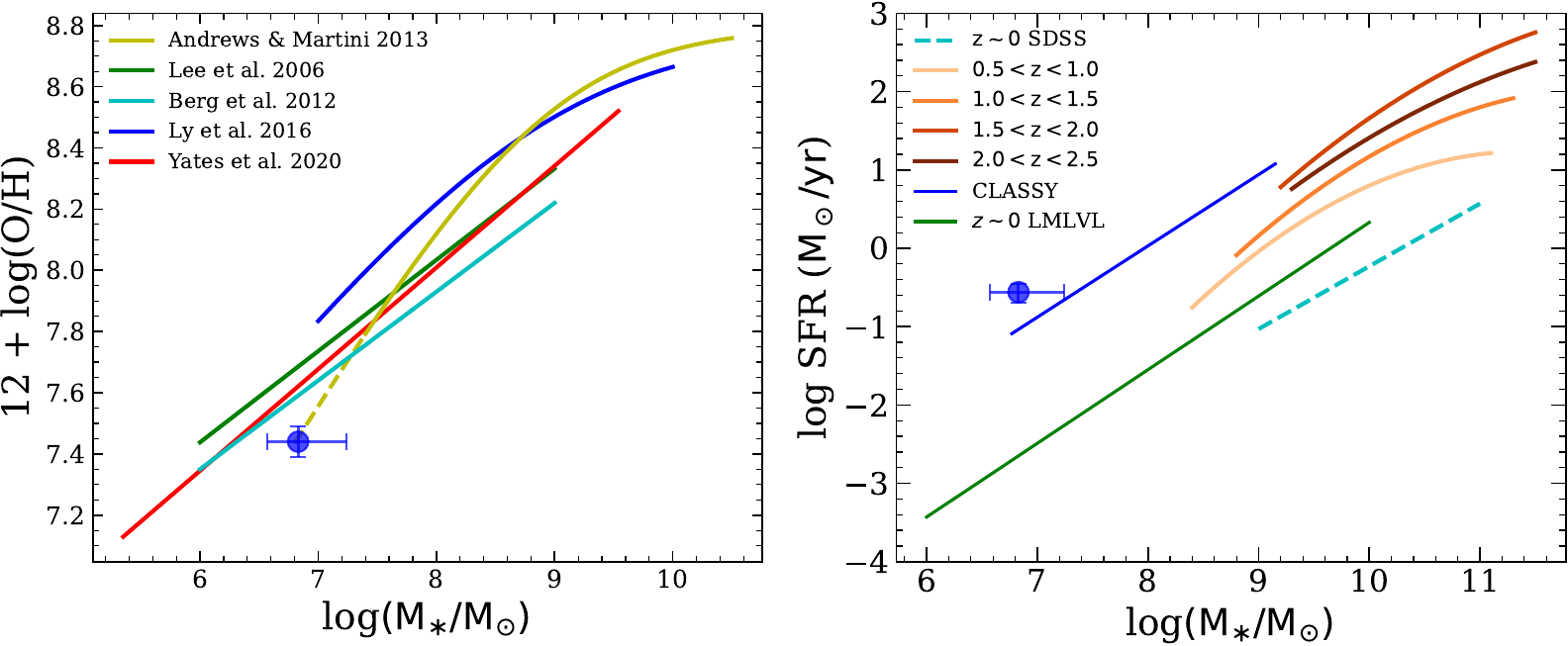}
\caption{{\em Left:} The direct-based MZRs. {Solid lines in various colors depict MZRs with direct-based metallicity measurements from multiple sources \citep{lee_extending_2006,berg_direct_2012,andrews_mass-metallicity_2013,Ly_2016,yates_present-day_2020}. The yellow dashed line illustrates the MZR extrapolation from \cite{andrews_mass-metallicity_2013} to the low-mass regime of J1044+0353. The total stellar mass of J1044+0353 is represented by a blue circle and extracted from \citetalias{berg_cos_2022}, with the median metallicity value and 1-$\sigma$ uncertainty ($\sigma$ = 0.05) from Figure \ref{fig:oh_map}. {\em Right:} Total star formation rates versus total stellar mass (SFMR). J1044+0353's position is indicated by a blue circle based on values from \citetalias{berg_cos_2022}, while the overall CLASSY sample trend is represented by a blue solid line \citepalias{berg_cos_2022}. The $z \sim 0.5 - 2.5$ SFMR trends from \citet{whitaker_constraining_2014} (orange lines) and $z \sim 0$ trends from \cite{chang_stellar_2015} (SDSS; cyan dashed line) and \cite{berg_direct_2012} (LMLVL; green solid line) are included for comparison.} \label{fig:mzr_sfr}}
\end{figure*}

\defcitealias{andrews_mass-metallicity_2013}{AM13}
We determine J1044+0353's position in the mass-metallicity plane using the median metallicity value presented in Figure \ref{fig:oh_map} and the total stellar mass ascertained by \citetalias{berg_cos_2022}. Due to the systematic offset between strong-line-based and direct-based MZRs \citep[e.g.,][]{lopez-sanchez_eliminating_2012}, we restrict our comparisons to the latter, as we employ the auroral line [O\,{\footnotesize III}] $\lambda 4363$ to derive the direct-based oxygen abundance in this study. As shown in Figure \ref{fig:mzr_sfr}, J1044+0353 is situated below most direct-based MZRs. However, it could fall within the MZR from \cite{andrews_mass-metallicity_2013} (hereafter AM13) upon extrapolating this relationship to the low-mass regime ($M_{\ast} \lesssim 10^7 M_{\odot}$). A key limitation of this argument is the omission of systematic differences in sample selections, stellar mass determination methods, and various approaches to measuring T$_{e}$, which result in slight deviations in slopes and normalizations of these MZR curves. Nevertheless, we can assert that J1044+0353 is more metal-depleted than most dwarf galaxies at this stellar mass.

To explain this offset from the MZR, prior research \citep[e.g.,][]{mannucci_fundamental_2010,curti_massmetallicity_2020} highlights the existence of a more general relation (the fundamental metallicity relation or FMR) among stellar mass, gas-phase metallicity, and SFR. These studies reveal a more pronounced secondary dependence of metallicity on SFR in the low-mass-end regime of the MZR, suggesting that low-metallicity galaxies exhibit higher SFRs at a fixed stellar mass. Given the steep low-mass slope of the MZR from \citetalias{andrews_mass-metallicity_2013} ($Z \propto M_{\ast}^{1/2}$; compared to other direct-based MZRs in Figure \ref{fig:mzr_sfr} with $Z \propto M_{\ast}^{1/3}$), the position of J1044+0353 is well-explained by extrapolating this MZR. \cite{Sanders_2021} ascribe this steep slope to a sample bias characterized by increasing SFR as $M_{\ast}$ declines, implying that this regime allows galaxies with varying SFRs to generate MZRs with distinct slopes.

This anti-correlation between metallicity and SFR suggests that J1044+0353's SFR is relatively higher than the typical SFRs at its stellar mass. J1044+0353's position in the SFR-$M_{\ast}$ (SFMR) plane verifies this point (shown in the right panel of Figure \ref{fig:mzr_sfr}). At its total stellar mass, it lies roughly 2 dex higher in SFR compared to the LMLVL samples in \cite{berg_direct_2012}. This point is also illustrated in \citetalias{berg_cos_2022}. The SFMR trend of the CLASSY (COS Legacy Archive Spectroscopic SurveY) galaxy sample, including J1044+0353, has a similar slope as the LMLVL trend at $z \sim 0$, but its SFR is more typical for galaxies around $z \sim 2$ \citep{whitaker_constraining_2014}.

\citet{mannucci_fundamental_2010} illustrate multiple approaches to explain the anti-correlation between metallicity and SFR discovered in the FMR. For example, galactic outflows are considered the primary mechanism for the dependence of abundances on mass and SFR in the local universe, especially for dwarf galaxies with shallower gravitational potential wells. 
If we extrapolate Figure 12 in \citetalias{andrews_mass-metallicity_2013} to this low-mass regime, we find this direct-based FMR significantly overestimates the metallicity value at this specific SFR and stellar mass (i.e., larger than the median metallicity value in Figure \ref{fig:oh_map} by $\sim$ 0.25 dex). 
This large deviation suggests outflow cannot be the sole mechanism that dominates in this galaxy.

Given the presence of localized chemical heterogeneities (Section \ref{subsubsec:oh_gradient}), we cannot exclude the possible impacts of mergers or interactions in J1044+0353's elevated SFR and diluted metallicity. Several studies find that merger-type galaxies are not consistent with the FMR \citep{ellison_galaxy_2008,bustamante_merger-induced_2018,bustamante_galaxy_2020}. Even though the FMR predicts a dilution of metallicity for weakly-interacting galaxies (i.e., projected separation $\gtrsim 50$ kpc), the measured metallicity offset for galaxies at the end stage of interaction is stronger than the FMR's prediction \citep{bustamante_galaxy_2020}. This metallicity offset or the SFR enhancement might be due to the induced radial inflow of low-metallicity gas (by mergers or interactions) from the outskirts to the galactic center \citep{rupke_galaxy_2010, rupke_gas-phase_2010, rich_integral_2012, torrey_metallicity_2012}. These studies indicate that J1044+0353 is an FMR outlier (i.e., the FMR is not universal) due to its interaction history.

\subsection{Spatial Variations in Metallicity} \label{subsec:spatial_variation_oh}

The conventional perspective that low-metallicity star-forming dwarf galaxies are chemically homogeneous is contested by several recent studies \citep[][and references therein]{Bresolin_2019,James_2020}. These studies propose that chemical inhomogeneities might arise in these systems under certain critical conditions, such as: (i) the presence of metal-enriched outflows stemming from supernovae and (ii) the induction of metal-poor gas inflows due to interactions and mergers.
We elaborate on each circumstance to account for the localized spatial variations in metallicity surrounding the post-starburst region (Section \ref{subsubsec:oh_gradient}).

\subsubsection{Potential Explanations} \label{subsubsec:potential_explanations}

Mergers or interactions could be the primary driver to localized chemical inhomogeneities. Numerical studies suggest that the induced inflows of less-enriched gas by mergers or strong interactions might lower the central metallicity \citep{rupke_galaxy_2010,rich_integral_2012,torrey_metallicity_2012}. Although these numerical simulations have different treatments of the chemical enrichment by star formation, gas consumption rate, and efficiency of galactic wind, all agree that nuclear metallicity is diluted after the first pericenter. This change in nuclear metallicity is not a monotonic function as the progress of a merger. Chemical enrichment and metal redistribution by galactic outflows would partially wipe out the dilution of metal-poor gas inflows. The metallicity dilution by mergers or interactions is also emphasized in observational studies \citep{rupke_gas-phase_2010,kewley_metallicity_2010,grossi_inverted_2020}. \citet{rich_integral_2012} observe that this dilution effect is prevalent across all stages of interaction, with the degree of dilution increasing from the weakly-interacting stage to the final coalesced stage for their luminous infrared galaxies, indicating that this effect is sensitive to the galaxy's interaction stage.

To specify a particular interaction stage for this galaxy, we first need to understand whether the chemical inhomogeneities are due to a strong encounter with a dark gas cloud or a dwarf-dwarf merger. The age difference ($\sim 20$ Myr) between the starburst and the post-starburst regions can constrain the parameters of these two scenarios. 

If the age difference is roughly the time it takes the gas cloud to pass through these two regions (separated by $\sim 1$ kpc), the velocity of this gas cloud is $\mathrm{1 \ kpc / 20 \ Myr} \approx 49 \ \mathrm{km/s}$. This value is a factor of three larger than the measured circular velocity of this galaxy ($15.6^{+3.0}_{-3.7} \ \mathrm{km/s}$) by \cite{xu_classy_2022}.

Alternatively, the age gradient might reflect the first and second pericenter of a dwarf-dwarf merger. If each of these two dwarf galaxies has a stellar mass of $10^6 \ M_{\odot}$ with a roughly $10^{9.5} \ M_{\odot}$ halo \citep[i.e., extrapolate the stellar mass-halo mass relation from \citet{Moster_2012} to the lower stellar-mass regime;][]{Wheeler_2019}, the estimated separation between them is $0.86$ kpc based on the assumption that they revolve in a circular orbit. This estimated separation is close to the projected distance between the starburst and the post-starburst regions.

Galactic outflows can also lead to spatial chemical variations. 
Observing the kinematic indicators of outflow gas in Knot C of the local blue compact dwarf (BCD) Haro 11, \citet{James_2013} attribute its depleted metallicity to a metal-enriched galactic outflow. 
\cite{wang_discovery_2019} discover two $z \sim 2$ dwarf galaxies with positive metallicity gradients due to strong outflows triggered by central starbursts. Intrinsically, the extent of metal enrichment probably depends on the fraction of supernovae-produced metals that mix with the ISM matter instead of being lost in metal-enriched galactic winds \citep{Sharda_Gradient_2021}.

\subsubsection{Analytical Chemical Evolution Model} \label{subsec:inflow_outflow}

To disentangle the roles of induced inflows (by merger or interaction) and galactic outflows played in this galaxy, we can apply the chemical evolution model of \cite{erb_model_2008} to find out the mass accretion factor, $f_{\mathrm{i}}  = \dot{M}_{\mathrm{acc}} / \mathrm{SFR}$, and the mass loading factor, $f_{\mathrm{o}} = \dot{M}_{\mathrm{out}} / \mathrm{SFR}$, needed to explain its metallicity and gas fraction $\mu$.


\begin{figure*}[htb]
\centering
\includegraphics[width=1.0\textwidth]{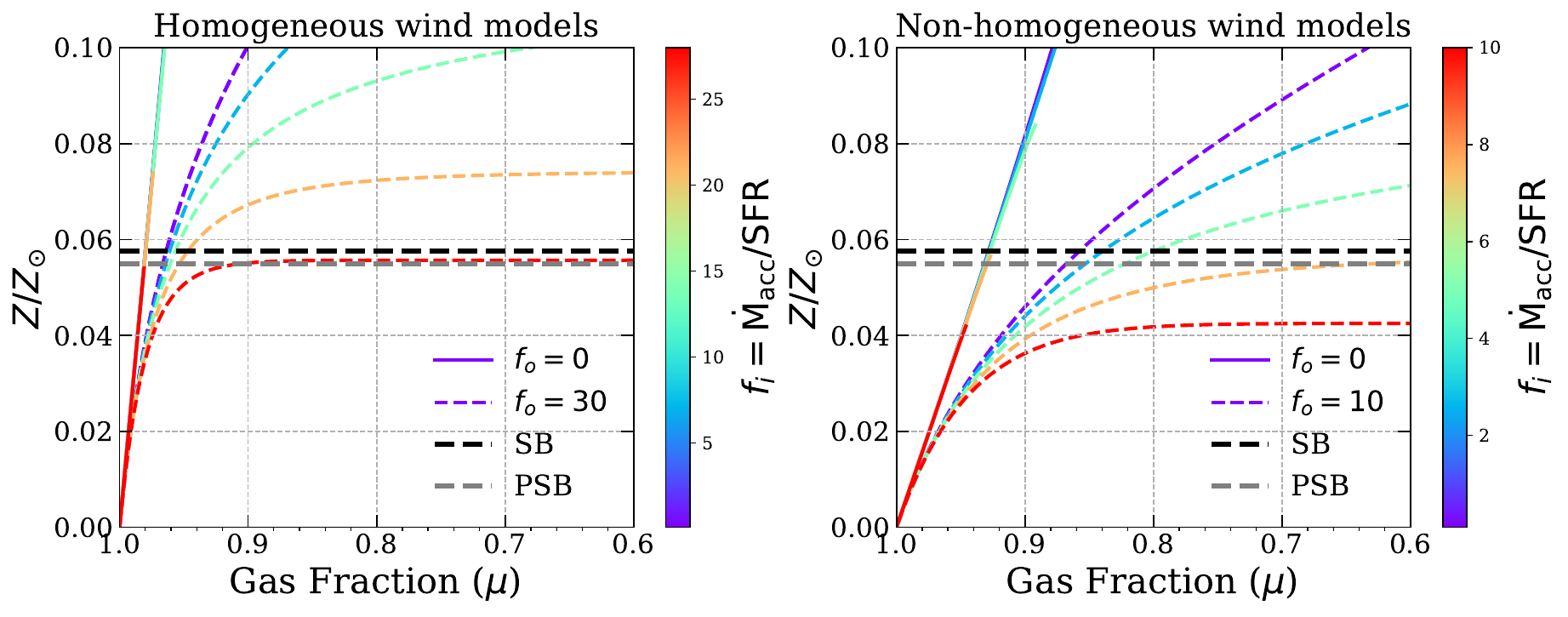}
\caption{Analytical evolution of metallicity with gas fraction from \cite{erb_model_2008} compared to the observed values of the starburst (SB) and the post-starburst (PSB) regions. {\em Left:} Homogeneous wind models such that $Z_{\mathrm{W}} = Z$. Evolution curves are shown at fixed outflow rates ($f_{\mathrm{o}} =$ 0 or 30), color-coded by inflow rate $f_{\mathrm{i}}$ ranging from 0 to 28. Inflow-only models (i.e., the solid lines) can explain the position of the starburst region ($\mu_{\mathrm{SB}} \gtrsim 0.98$), which is consistent with Figure \ref{fig:velocity_map} that $v_p$ is low around this region. The property of the post-starburst region ($0.70 \lesssim \mu_{\mathrm{PSB}} \lesssim 0.85$) can only be reproduced by a high-$f_{\mathrm{i}}$-high-$f_{\mathrm{o}}$ model (e.g., $f_{\mathrm{i}} = 28$ and $f_{\mathrm{o}} = 30$), which is unrealistic.
{\em Right:} Non-homogeneous wind models such that $Z_{\mathrm{W}}$ is defined in Equation \ref{4.1}. Evolution curves are shown at fixed outflow rates ($f_{\mathrm{o}} =$ 0 or 10), color-coded by inflow rate $f_{\mathrm{i}}$ ranging from 0 to 10. Compared to the homogeneous model, a reasonable range of $f_{\mathrm{i}}$ (i.e., $2 \lesssim f_{\mathrm{i}} \lesssim 8$) at a physical outflow rate $f_{\mathrm{o}} = 10$ can explain the position of the post-starburst region. In each panel, the horizontal black (or grey) dashed line denotes the gas-phase metallicity for the starburst (or post-starburst) region. \label{fig:erb08}} 
\end{figure*}

We first convert the SFR surface densities of the starburst and the post-starburst region to the gas surface densities by inverting the Kennicutt-Schmidt law \citep{Kennicutt_Jr__1998}. This approach delineates rough constraints on gas fractions for both regions: $\mu_{\mathrm{SB}} \gtrsim 0.98$ and $0.70 \lesssim \mu_{\mathrm{PSB}} \lesssim 0.85$, with the latter representing the stellar-weighted gas fraction in PSB. Assuming a return fraction of $R = 0.45$ and a true stellar yield of $y = 0.038$ for the \cite{chabrier_galactic_2003} IMF and at metallicity $ \sim 0.08 Z_{\odot}$ \citep{Vincenzo_2016}, we then calculate the evolution curves of metallicity with gas fraction at fixed $f_{\mathrm{o}}$ (0 or 30) using Equation (11) and (13) from \citet{erb_model_2008}. This model assumes a homogeneous wind that the outflowing gas has the same metallicity as the gas in the ISM. The left panel of Figure \ref{fig:erb08} shows inflow-only models (i.e., the solid lines) can explain the position of the starburst region, which is consistent with Figure \ref{fig:velocity_map} that $v_p$ is low around this region. These models converge to a single curve in the high-$\mu$ regime, so it is hard for us to assign an exact value of $f_{\mathrm{i}}$ for the starburst region. 

The property of the post-starburst region can only be reproduced by a high-$f_{\mathrm{i}}$-high-$f_{\mathrm{o}}$ model (e.g., $f_{\mathrm{i}} = 28$ and $f_{\mathrm{o}} = 30$; the red dashed line in the left panel of Figure \ref{fig:erb08}) given the constraint on its gas fraction. This unrealistic inflow rate questions the validity of the assumption of homogeneous wind in this dwarf galaxy. Our metallicity measurement also suggests metals ejected in the outflowing gas are not well-mixed with the ISM, which is compatible with previous studies \citep[e.g.,][]{Chisholm_2018}. To relax this assumption, we parametrize the wind metallicity $Z_{\mathrm{W}}$ as Equation (41) of \cite{Forbes_2019} in Equation \ref{4.1}. 
\begin{equation}
\label{4.1} Z_\mathrm{W} = Z + \xi \frac{y}{\mathrm{max(\mu, R)}} 
\end{equation}
We use a fiducial value of $\xi = 0.4$ ($0 \leq \xi \leq 1$) that is appropriate for $\mathrm{log(M_{\ast}/M_{\odot})} \sim 8$ galaxies in \citet{curti_massmetallicity_2020} \citep{sharda2021origin}. $\xi = 0$ corresponds to the homogeneous wind model, whereas $\xi = 1$ represents the scenario that all supernova ejecta are lost in winds.  

Following this non-homogeneous wind model (see the right panel of Figure \ref{fig:erb08}), we find that a reasonable range of $f_{\mathrm{i}}$ (i.e., $2 \lesssim f_{\mathrm{i}} \lesssim 8$) at a physical outflow rate $f_{\mathrm{o}} = 10$ can explain the position of the post-starburst region. Therefore, as long as the wind from the post-starburst region is metal-enriched, we do not require unphysically high outflow and inflow rates to reproduce our observations. We should also keep in mind this finding is based on the fiducial value of $\xi$ that is poorly constrained by both theory and observations, and it is degenerate with $f_{\mathrm{i}}$ and $f_{\mathrm{o}}$. As a result, accurate measurements of $\xi$ and $\mu$ would help us break the degeneracy among these physical quantities.

\section{Conclusion} \label{sec:conclusion}
The dwarf galaxy J1044+0353 provides an ideal test-bed for understanding high-redshift galaxies during the Epoch of Reionization. Previous studies have drawn attention to the extreme physical properties (e.g., low metallicity, high ionization, and high specific star formation rate) of the starburst region. In this paper, we used integral field spectroscopy to map the metallicity and gas kinematics over spatial scales much larger than the compact starburst. We also examined stellar spectral features and their spatial variations. Our results provide insight into the origin of the starburst and its feedback on the surrounding interstellar medium. These gas flows strongly influence the galaxy's chemical evolution during this phase of rapid mass assembly.

We summarize our specific results from this work as follows:
\begin{enumerate}
  \item We discovered a post-starburst region which we divided into partially resolved associations of stars:  MEN, NEN, and FEN. We found significant differences among the emission-line and absorption-line equivalent widths of these regions. We proposed a composite SFH using BEAGLE to constrain their stellar ages. Specifically, we constructed EW grids (i.e., x-axis: EW of the emission portion of a line profile; y-axis: EW of the absorption portion of a line profile) for the Balmer lines $\mathrm{H\gamma}$ and $\mathrm{H\delta}$. Based on each region's position in the EW grids, we estimate the log(Age/year) of each stellar association to be $6.480 \pm 0.005$ (SB), $7.2 \pm 0.1$ (MEN $\&$ NEN), and $7.30 \pm 0.05$ (FEN), respectively. Our age estimation for the starburst is consistent with the one derived by  \cite{Olivier_2022} from the FUV spectrum.
  The time interval between the post-starburst and the starburst may reflect the propagation of a disturbance across the galaxy, e.g., an encounter with a gas cloud or a mini-halo. Alternatively, the age gradient might reflect the first and second passage of a dwarf-dwarf merger.
  
  \item We also used the mass-to-light ratios of mock galaxies to determine the stellar mass of each region. The log($M_\ast$/$M_\odot$) was measured at $5.82^{+ 0.39}_{- 0.12}$, $6.26^{+ 0.05}_{- 0.18}$, $6.08^{+ 0.05}_{- 0.18}$ and $5.57^{+ 0.10}_{- 0.04}$ for SB, MEN, NEN, and FEN, respectively. 

  \item The intrinsic extinction $A_V$ of each stellar association is determined from the Balmer decrement. 
  SB ($0.20^{+0.10}_{-0.10}$), FEN ($0.20^{+0.04}_{-0.04}$), and NEN ($0.21^{+0.13}_{-0.12}$) are found to have similar dust attenuation. 
  However, the intrinsic extinction is significant in MEN ($0.56^{+0.09}_{-0.09}$), implying a dust lane, perhaps a relic of a galaxy merger. 

  \item Using the temperature sensitive auroral line [O\,{\footnotesize III}] $\lambda 4363$, we derived the ``direct'' gas-phase oxygen abundance of $7.45 \pm 0.01$ (SB), $7.37 \pm 0.03$ (MEN), $7.52 \pm 0.03$ (NEN),  and $7.43 \pm 0.03$ (FEN) in terms of 12 +log(O/H) respectively. The stellar-mass-weighted metallicity of  the post-starburst region (12 +log(O/H) = $7.43 \pm 0.02$) is comparable to that of the starburst region, which indicates metal-enriched wind. There were subtle differences between our ``direct'' method and the ones used in \citetalias{berg_carbon_2016} and \citetalias{berg_characterizing_2021} to measure metallicity, including the calibration between T[O\,{\footnotesize III}] and T[O\,{\footnotesize II}], the electron density diagnostics, the Balmer lines chose to determine the intrinsic reddening value, the intrinsic reddening law, and the ionization-zone model (two or four ionization zones). The deviations of metallicity in the starburst region are $\lesssim 0.01$ dex among our work and previous studies. 

  \item Applying an adaptive binning algorithm to the strongest emission line [O\,{\footnotesize III}] $\lambda 5007$, we created a line-of-sight velocity map. We discovered a strong velocity gradient along the galaxy's minor axis, up to 50 km/s over 1.5 kpc. If this gradient represents rotation, the stellar system would have the shape of a prolate spheroid rotating about its long axis, a scenario that is highly unlikely. 
  We argue that the galaxy is better described as an oblate spheroid, essentially a gas disk viewed at a fairly high inclination; then, we explain the minor-axis velocity gradient by a bipolar outflow erupting perpendicular to the gas disk. A closer examination of the velocity field supports the outflow interpretation. The outflow cones converge at a vertex near FEN, the oldest stellar association, that marks the location where the most supernova explosions have occurred. In addition,
  the regions with high absolute peculiar velocities also have the largest line widths. This correspondence supports the outflow interpretation because the skew of the line profiles matches expectations for an inclined cone. The outflow speed is higher than the estimated escape velocity for such a low-mass galaxy, suggesting a possible escape from the gravitational potential.

  \item The asymmetric velocity field around the starburst region is notable, indicating the systematic velocity offset between the starburst's southern regions and the entire galaxy. This velocity offset reveals the existence of induced inflows fueling the starburst region during mergers or interactions, a hypothesis awaiting future H\,{\footnotesize I} mapping for confirmation.

  \item For its stellar mass, the metallicity of J1044+0353 lies below most MZRs in the literature. Its diluted  metallicity is qualitatively consistent with its elevated SFR, and the galaxy lies on the extrapolation of the MZR defined by \citep{andrews_mass-metallicity_2013}. From the direct-based metallicity map, J1044+0353 is considered chemically homogeneous due to its relatively small IQR and $\sigma_{\mathrm{Z}}$. However, we observe a relatively large scatter of metallicities within the post-starburst region compared to the starburst region, with the largest spread of approximately 0.3 dex between individual regions. This increased scatter in the post-starburst region suggests an induced inflow of low-metallicity gas, potentially caused by mergers or interactions.
  
 \vspace{-0.5em}
  \item
  To quantify the outflow and the induced inflow rates, we first applied the \citet{erb_model_2008} chemical evolution model, which includes a pristine-gas inflow and a homogeneous wind. For the starburst metallicity, we found that inflow-only models require a near-unity gas fraction, consistent with the high SFR surface density and Kennicutt-Schmidt law; the inflow rate is not well constrained. Models with outflows have lower gas fractions when they reach the SB metallicity. Quantitatively, however, even high mass loss rates, $f_\mathrm{o} = \dot{M}_{\mathrm{out}}/ \mathrm{SFR} = 30$, are not ruled out;  for inflow rates $f_\mathrm{i} = \dot{M}_{\mathrm{acc}}/ \mathrm{SFR} \lesssim 5$, the required gas fraction still exceeds 96\%, which is roughly within observational constraints. 

   \vspace{-0.5em}
  \item
    Because the gas fraction is arguably lower in the post-starburst region, homogeneous models require enormous gas outflow and inflow rates to reduce the gas fraction below $\sim 95\%$ at its gas-phase metallicity. One plausible solution is that the metals in the outflowing gas are not well-mixed with the gas in the ISM. We demonstrated that a chemical evolution model with a metal-enriched wind reaches the post-starburst metallicity at a significantly lower gas fraction. We prefer this metal-enriched wind model for the chemical evolution of J1044+0353 because it requires reasonable outflow and inflow rates in the post-starburst region.

\end{enumerate}

Our analysis provides detailed insight into the physical processes that shape early galaxy assembly.  If J1044+0353 were at high redshift, only the compact starburst would be detected.  While the young age of the starburst would suggest relatively little supernova feedback, our study shows a different picture.  The duration of the starburst is extended over roughly a dynamical timescale, long enough for supernova feedback to create holes in the surrounding neutral gas reservoir before all the O stars die off, causing ionizing photons to escape and reionize the intergalactic gas.

Future KCRM or archived
VLT/MUSE observations could cover the $\mathrm{H\alpha}$ line. A $\mathrm{H\alpha/\mathrm{H\beta}}$ map would substantially improve the accuracy of the reddening map and offer additional constraints on the optically thin regions. These regions, corresponding to holes in the H\,{\footnotesize I} distribution, might shed light on the escape direction of ionizing photons.


\software{{\tt\string PyNeb} \citep[v1.1.14,][]{pyneb,pyneb_2015},
{\tt\string LMFIT} \citep{lmfit}, BEAGLE \citep{chevallard_modelling_2016},
{\tt\string Photutils} \citep{larry_bradley_2022_6825092}, {\tt\string Petrofit} \citep{Geda_2022}}

\begin{acknowledgments}
Z.P. would like to acknowledge Giovanni Cresci, Xin Wang, Zuyi Chen, and Zhuyun Zhuang for helpful discussions and Christophe Morisset for technical assistance in {\tt\string PyNeb}. The authors are grateful to the anonymous referee for constructive suggestions. We gratefully acknowledge support for this work by NSF grant AST-1817125.
\end{acknowledgments}

\bibliography{sample631}{}

\begin{thebibliography}{}
\expandafter\ifx\csname natexlab\endcsname\relax\def\natexlab#1{#1}\fi
\providecommand{\url}[1]{\href{#1}{#1}}
\providecommand{\dodoi}[1]{doi:~\href{http://doi.org/#1}{\nolinkurl{#1}}}
\providecommand{\doeprint}[1]{\href{http://ascl.net/#1}{\nolinkurl{http://ascl.net/#1}}}
\providecommand{\doarXiv}[1]{\href{https://arxiv.org/abs/#1}{\nolinkurl{https://arxiv.org/abs/#1}}}

\bibitem[{Andrews \& Martini(2013)}]{andrews_mass-metallicity_2013}
Andrews, B.~H., \& Martini, P. 2013, The Astrophysical Journal, 765, 140,
  \dodoi{10.1088/0004-637x/765/2/140}

\bibitem[{Arellano-C{\'{o}}rdova \&
  Rodr{\'{i}}guez(2020)}]{arellano-cordova_ten_2020}
Arellano-C{\'{o}}rdova, K.~Z., \& Rodr{\'{i}}guez, M. 2020, Monthly Notices of
  the Royal Astronomical Society, 497, 672, \dodoi{10.1093/mnras/staa1759}

\bibitem[{Arellano-C{\'{o}}rdova {et~al.}(2022)Arellano-C{\'{o}}rdova,
  Mingozzi, Berg, James, Rogers, Aloisi, Amorín, Brinchmann, Charlot,
  Chisholm, Heckman, Dubón, Hayes, Hernandez, Jones, Kumari, Leitherer,
  Martin, Nanayakkara, Pogge, Sanders, Senchyna, Skillman, Stark, Wofford, \&
  Xu}]{arellano-cordova_classy_2022}
Arellano-C{\'{o}}rdova, K.~Z., Mingozzi, M., Berg, D.~A., {et~al.} 2022,
  {CLASSY} V: The impact of aperture effects on the inferred nebular properties
  of local star-forming galaxies,  {arXiv}

\bibitem[{{Asplund} {et~al.}(2021){Asplund}, {Amarsi}, \&
  {Grevesse}}]{Asplund_2021}
{Asplund}, M., {Amarsi}, A.~M., \& {Grevesse}, N. 2021, \aap, 653, A141,
  \dodoi{10.1051/0004-6361/202140445}

\bibitem[{{Barnes} \& {Hernquist}(1996)}]{BH_96}
{Barnes}, J.~E., \& {Hernquist}, L. 1996, \apj, 471, 115,
  \dodoi{10.1086/177957}

\bibitem[{{Barnes} \& {Hernquist}(1991)}]{BH_91}
{Barnes}, J.~E., \& {Hernquist}, L.~E. 1991, \apjl, 370, L65,
  \dodoi{10.1086/185978}

\bibitem[{Berg {et~al.}(2021)Berg, Chisholm, Erb, Skillman, Pogge, \&
  Olivier}]{berg_characterizing_2021}
Berg, D.~A., Chisholm, J., Erb, D.~K., {et~al.} 2021, The Astrophysical
  Journal, 922, 170, \dodoi{10.3847/1538-4357/ac141b}

\bibitem[{Berg {et~al.}(2016)Berg, Skillman, Henry, Erb, \&
  Carigi}]{berg_carbon_2016}
Berg, D.~A., Skillman, E.~D., Henry, R. B.~C., Erb, D.~K., \& Carigi, L. 2016,
  The Astrophysical Journal, 827, 126, \dodoi{10.3847/0004-637x/827/2/126}

\bibitem[{Berg {et~al.}(2012)Berg, Skillman, Marble, van Zee, Engelbracht, Lee,
  Kennicutt, Calzetti, Dale, \& Johnson}]{berg_direct_2012}
Berg, D.~A., Skillman, E.~D., Marble, A.~R., {et~al.} 2012, The Astrophysical
  Journal, 754, 98, \dodoi{10.1088/0004-637x/754/2/98}

\bibitem[{Berg {et~al.}(2022)Berg, James, King, McDonald, Chen, Chisholm,
  Heckman, Martin, Stark, Aloisi, Amor{\'{\i}}n, Arellano-C{\'{o}}rdova,
  Bayliss, Bordoloi, Brinchmann, Charlot, Chevallard, Clark, Erb, Feltre,
  Gronke, Hayes, Henry, Hernandez, Jaskot, Jones, Kewley, Kumari, Leitherer,
  Llerena, Maseda, Mingozzi, Nanayakkara, Ouchi, Plat, Pogge, Ravindranath,
  Rigby, Sanders, Scarlata, Senchyna, Skillman, Steidel, Strom, Sugahara,
  Wilkins, Wofford, \& Xu}]{berg_cos_2022}
Berg, D.~A., James, B.~L., King, T., {et~al.} 2022, The Astrophysical Journal
  Supplement Series, 261, 31, \dodoi{10.3847/1538-4365/ac6c03}

\bibitem[{{Blum} {et~al.}(2016){Blum}, {Burleigh}, {Dey}, {Schlegel},
  {Meisner}, {Levi}, {Myers}, {Lang}, {Moustakas}, {Patej}, {Valdes}, {Kneib},
  {Huanyuan}, {Nord}, {Olsen}, {Delubac}, {Saha}, {James}, {Walker}, \& {DECaLS
  Team}}]{decals_16}
{Blum}, R.~D., {Burleigh}, K., {Dey}, A., {et~al.} 2016, in American
  Astronomical Society Meeting Abstracts, Vol. 228, American Astronomical
  Society Meeting Abstracts \#228, 317.01

\bibitem[{Boissier \& Prantzos(2000)}]{boissier_chemo-spectrophotometric_2000}
Boissier, S., \& Prantzos, N. 2000, Monthly Notices of the Royal Astronomical
  Society, 312, 398, \dodoi{10.1046/j.1365-8711.2000.03133.x}

\bibitem[{Bradley {et~al.}(2022)Bradley, Sipőcz, Robitaille, Tollerud,
  Vinícius, Deil, Barbary, Wilson, Busko, Donath, Günther, Cara, Lim,
  Meßlinger, Conseil, Bostroem, Droettboom, Bray, Bratholm, Barentsen, Craig,
  Rathi, Pascual, Perren, Georgiev, de~Val-Borro, Kerzendorf, Bach, Quint, \&
  Souchereau}]{larry_bradley_2022_6825092}
Bradley, L., Sipőcz, B., Robitaille, T., {et~al.} 2022, astropy/photutils:
  1.5.0, 1.5.0,  Zenodo, \dodoi{10.5281/zenodo.6825092}

\bibitem[{{Bresolin}(2019)}]{Bresolin_2019}
{Bresolin}, F. 2019, \mnras, 488, 3826, \dodoi{10.1093/mnras/stz1947}

\bibitem[{Bustamante {et~al.}(2020)Bustamante, Ellison, Patton, \&
  Sparre}]{bustamante_galaxy_2020}
Bustamante, S., Ellison, S.~L., Patton, D.~R., \& Sparre, M. 2020, Monthly
  Notices of the Royal Astronomical Society, 494, 3469,
  \dodoi{10.1093/mnras/staa1025}

\bibitem[{Bustamante {et~al.}(2018)Bustamante, Sparre, Springel, \&
  Grand}]{bustamante_merger-induced_2018}
Bustamante, S., Sparre, M., Springel, V., \& Grand, R. J.~J. 2018, Monthly
  Notices of the Royal Astronomical Society, 479, 3381,
  \dodoi{10.1093/mnras/sty1692}

\bibitem[{Campbell {et~al.}(1986)Campbell, Terlevich, \&
  Melnick}]{campbell_stellar_1986}
Campbell, A., Terlevich, R., \& Melnick, J. 1986, Monthly Notices of the Royal
  Astronomical Society, 223, 811, \dodoi{10.1093/mnras/223.4.811}

\bibitem[{Cappellari \& Copin(2003)}]{cappellari_adaptive_2003}
Cappellari, M., \& Copin, Y. 2003, Monthly Notices of the Royal Astronomical
  Society, 342, 345, \dodoi{10.1046/j.1365-8711.2003.06541.x}

\bibitem[{{Cardelli} {et~al.}(1989){Cardelli}, {Clayton}, \&
  {Mathis}}]{cardelli_1989}
{Cardelli}, J.~A., {Clayton}, G.~C., \& {Mathis}, J.~S. 1989, \apj, 345, 245,
  \dodoi{10.1086/167900}

\bibitem[{{Carr} {et~al.}(2022){Carr}, {Michel-Dansac}, {Blaizot}, {Scarlata},
  {Henry}, \& {Verhamme}}]{Carr_2022}
{Carr}, C., {Michel-Dansac}, L., {Blaizot}, J., {et~al.} 2022, arXiv e-prints,
  arXiv:2209.14473, \dodoi{10.48550/arXiv.2209.14473}

\bibitem[{Chabrier(2003)}]{chabrier_galactic_2003}
Chabrier, G. 2003, \pasp, 115, 763, \dodoi{10.1086/376392}

\bibitem[{Chang {et~al.}(2015)Chang, van~der Wel, da~Cunha, \&
  Rix}]{chang_stellar_2015}
Chang, Y.-Y., van~der Wel, A., da~Cunha, E., \& Rix, H.-W. 2015, The
  Astrophysical Journal Supplement Series, 219, 8,
  \dodoi{10.1088/0067-0049/219/1/8}

\bibitem[{Charlot \& Fall(2000)}]{charlot_simple_2000}
Charlot, S., \& Fall, S.~M. 2000, The Astrophysical Journal, 539, 718,
  \dodoi{10.1086/309250}

\bibitem[{Chevallard \& Charlot(2016)}]{chevallard_modelling_2016}
Chevallard, J., \& Charlot, S. 2016, Monthly Notices of the Royal Astronomical
  Society, 462, 1415, \dodoi{10.1093/mnras/stw1756}

\bibitem[{Chevallard {et~al.}(2018)Chevallard, Charlot, Senchyna, Stark,
  Vidal-García, Feltre, Gutkin, Jones, Mainali, \&
  Wofford}]{chevallard_physical_2018}
Chevallard, J., Charlot, S., Senchyna, P., {et~al.} 2018, Monthly Notices of
  the Royal Astronomical Society, 479, 3264, \dodoi{10.1093/mnras/sty1461}

\bibitem[{{Chisholm} {et~al.}(2019){Chisholm}, {Rigby}, {Bayliss}, {Berg},
  {Dahle}, {Gladders}, \& {Sharon}}]{Chisholm_2019}
{Chisholm}, J., {Rigby}, J.~R., {Bayliss}, M., {et~al.} 2019, \apj, 882, 182,
  \dodoi{10.3847/1538-4357/ab3104}

\bibitem[{{Chisholm} {et~al.}(2018){Chisholm}, {Tremonti}, \&
  {Leitherer}}]{Chisholm_2018}
{Chisholm}, J., {Tremonti}, C., \& {Leitherer}, C. 2018, \mnras, 481, 1690,
  \dodoi{10.1093/mnras/sty2380}

\bibitem[{Cresci {et~al.}(2010)Cresci, Mannucci, Maiolino, Marconi, Gnerucci,
  \& Magrini}]{cresci_gas_2010}
Cresci, G., Mannucci, F., Maiolino, R., {et~al.} 2010, Nature, 467, 811,
  \dodoi{10.1038/nature09451}

\bibitem[{Curti {et~al.}(2020)Curti, Mannucci, Cresci, \&
  Maiolino}]{curti_massmetallicity_2020}
Curti, M., Mannucci, F., Cresci, G., \& Maiolino, R. 2020, Monthly Notices of
  the Royal Astronomical Society, 491, 944, \dodoi{10.1093/mnras/stz2910}

\bibitem[{Davé {et~al.}(2011)Davé, Finlator, \& Oppenheimer}]{dave_2011}
Davé, R., Finlator, K., \& Oppenheimer, B.~D. 2011, \mnras, 416, 1354,
  \dodoi{10.1111/j.1365-2966.2011.19132.x}

\bibitem[{{De Rossi} {et~al.}(2017){De Rossi}, {Bower}, {Font}, {Schaye}, \&
  {Theuns}}]{Rossi_2017}
{De Rossi}, M.~E., {Bower}, R.~G., {Font}, A.~S., {Schaye}, J., \& {Theuns}, T.
  2017, \mnras, 472, 3354, \dodoi{10.1093/mnras/stx2158}

\bibitem[{{Dey} {et~al.}(2019){Dey}, {Schlegel}, {Lang}, {Blum}, {Burleigh},
  {Fan}, {Findlay}, {Finkbeiner}, {Herrera}, {Juneau}, {Landriau}, {Levi},
  {McGreer}, {Meisner}, {Myers}, {Moustakas}, {Nugent}, {Patej}, {Schlafly},
  {Walker}, {Valdes}, {Weaver}, {Y{\`e}che}, {Zou}, {Zhou}, {Abareshi},
  {Abbott}, {Abolfathi}, {Aguilera}, {Alam}, {Allen}, {Alvarez}, {Annis},
  {Ansarinejad}, {Aubert}, {Beechert}, {Bell}, {BenZvi}, {Beutler}, {Bielby},
  {Bolton}, {Brice{\~n}o}, {Buckley-Geer}, {Butler}, {Calamida}, {Carlberg},
  {Carter}, {Casas}, {Castander}, {Choi}, {Comparat}, {Cukanovaite}, {Delubac},
  {DeVries}, {Dey}, {Dhungana}, {Dickinson}, {Ding}, {Donaldson}, {Duan},
  {Duckworth}, {Eftekharzadeh}, {Eisenstein}, {Etourneau}, {Fagrelius},
  {Farihi}, {Fitzpatrick}, {Font-Ribera}, {Fulmer}, {G{\"a}nsicke},
  {Gaztanaga}, {George}, {Gerdes}, {Gontcho}, {Gorgoni}, {Green}, {Guy},
  {Harmer}, {Hernandez}, {Honscheid}, {Huang}, {James}, {Jannuzi}, {Jiang},
  {Joyce}, {Karcher}, {Karkar}, {Kehoe}, {Kneib}, {Kueter-Young}, {Lan},
  {Lauer}, {Le Guillou}, {Le Van Suu}, {Lee}, {Lesser}, {Perreault Levasseur},
  {Li}, {Mann}, {Marshall}, {Mart{\'\i}nez-V{\'a}zquez}, {Martini}, {du Mas des
  Bourboux}, {McManus}, {Meier}, {M{\'e}nard}, {Metcalfe},
  {Mu{\~n}oz-Guti{\'e}rrez}, {Najita}, {Napier}, {Narayan}, {Newman}, {Nie},
  {Nord}, {Norman}, {Olsen}, {Paat}, {Palanque-Delabrouille}, {Peng},
  {Poppett}, {Poremba}, {Prakash}, {Rabinowitz}, {Raichoor}, {Rezaie},
  {Robertson}, {Roe}, {Ross}, {Ross}, {Rudnick}, {Safonova}, {Saha},
  {S{\'a}nchez}, {Savary}, {Schweiker}, {Scott}, {Seo}, {Shan}, {Silva},
  {Slepian}, {Soto}, {Sprayberry}, {Staten}, {Stillman}, {Stupak}, {Summers},
  {Sien Tie}, {Tirado}, {Vargas-Maga{\~n}a}, {Vivas}, {Wechsler}, {Williams},
  {Yang}, {Yang}, {Yapici}, {Zaritsky}, {Zenteno}, {Zhang}, {Zhang}, {Zhou}, \&
  {Zhou}}]{desi_2019}
{Dey}, A., {Schlegel}, D.~J., {Lang}, D., {et~al.} 2019, \aj, 157, 168,
  \dodoi{10.3847/1538-3881/ab089d}

\bibitem[{Ellison {et~al.}(2008)Ellison, Patton, Simard, \&
  {McConnachie}}]{ellison_galaxy_2008}
Ellison, S.~L., Patton, D.~R., Simard, L., \& {McConnachie}, A.~W. 2008, The
  Astronomical Journal, 135, 1877, \dodoi{10.1088/0004-6256/135/5/1877}

\bibitem[{Erb(2008)}]{erb_model_2008}
Erb, D.~K. 2008, The Astrophysical Journal, 674, 151, \dodoi{10.1086/524727}

\bibitem[{{Fern{\'a}ndez-Arenas} {et~al.}(2023){Fern{\'a}ndez-Arenas},
  {Carrasco}, {Terlevich}, {Terlevich}, {Amor{\'\i}n}, {Bresolin},
  {Ch{\'a}vez}, {Gonz{\'a}lez-Mor{\'a}n}, {Rosa-Gonz{\'a}lez}, {Mayya}, {Vega},
  {Zaragoza-Cardiel}, {M{\'e}ndez-Abreu}, {Izazaga-P{\'e}rez}, {Gil de Paz},
  {Gallego}, {Iglesias-P{\'a}ramo}, {Garc{\'\i}a-Vargas}, {G{\'o}mez-Alvarez},
  {Castillo-Morales}, {Cardiel}, {Pascual}, \& {P{\'e}rez-Calpena}}]{FD_2023}
{Fern{\'a}ndez-Arenas}, D., {Carrasco}, E., {Terlevich}, R., {et~al.} 2023,
  \mnras, 519, 4221, \dodoi{10.1093/mnras/stac3309}

\bibitem[{Fitzpatrick(1999)}]{Fitzpatrick_1999}
Fitzpatrick, E.~L. 1999, Publications of the Astronomical Society of the
  Pacific, 111, 63, \dodoi{10.1086/316293}

\bibitem[{{Forbes} {et~al.}(2019){Forbes}, {Krumholz}, \&
  {Speagle}}]{Forbes_2019}
{Forbes}, J.~C., {Krumholz}, M.~R., \& {Speagle}, J.~S. 2019, \mnras, 487,
  3581, \dodoi{10.1093/mnras/stz1473}

\bibitem[{Garnett(1992)}]{garnett_electron_1992}
Garnett, D.~R. 1992, The Astronomical Journal, 103, 1330,
  \dodoi{10.1086/116146}

\bibitem[{Geda {et~al.}(2022)Geda, Crawford, Hunt, Bershady, Tollerud, \&
  Randriamampandry}]{Geda_2022}
Geda, R., Crawford, S.~M., Hunt, L., {et~al.} 2022, The Astronomical Journal,
  163, 202, \dodoi{10.3847/1538-3881/ac5908}

\bibitem[{Gordon {et~al.}(2003)Gordon, Clayton, Misselt, Landolt, \&
  Wolff}]{gordon_2003}
Gordon, K.~D., Clayton, G.~C., Misselt, K.~A., Landolt, A.~U., \& Wolff, M.~J.
  2003, The Astrophysical Journal, 594, 279, \dodoi{10.1086/376774}

\bibitem[{{Grossi} {et~al.}(2020){Grossi}, {Garc{\'\i}a-Benito}, {Cortesi},
  {Gon{\c{c}}alves}, {Gon{\c{c}}alves}, {Lopes}, {Men{\'e}ndez-Delmestre}, \&
  {Telles}}]{grossi_inverted_2020}
{Grossi}, M., {Garc{\'\i}a-Benito}, R., {Cortesi}, A., {et~al.} 2020, \mnras,
  498, 1939, \dodoi{10.1093/mnras/staa2382}

\bibitem[{{Haurberg} {et~al.}(2013){Haurberg}, {Rosenberg}, \&
  {Salzer}}]{Haurberg_2013}
{Haurberg}, N.~C., {Rosenberg}, J., \& {Salzer}, J.~J. 2013, \apj, 765, 66,
  \dodoi{10.1088/0004-637X/765/1/66}

\bibitem[{Heckman {et~al.}(1990)Heckman, Armus, \& Miley}]{heckman_nature_1990}
Heckman, T.~M., Armus, L., \& Miley, G.~K. 1990, {ApJS}, 74, 833,
  \dodoi{10.1086/191522}

\bibitem[{{Ho} {et~al.}(2014){Ho}, {Kewley}, {Dopita}, {Medling}, {Allen},
  {Bland-Hawthorn}, {Bloom}, {Bryant}, {Croom}, {Fogarty}, {Goodwin}, {Green},
  {Konstantopoulos}, {Lawrence}, {L{\'o}pez-S{\'a}nchez}, {Owers}, {Richards},
  \& {Sharp}}]{Ho_2014}
{Ho}, I.~T., {Kewley}, L.~J., {Dopita}, M.~A., {et~al.} 2014, \mnras, 444,
  3894, \dodoi{10.1093/mnras/stu1653}

\bibitem[{Hwang {et~al.}(2019)Hwang, Barrera-Ballesteros, Heckman, Rowlands,
  Lin, Rodriguez-Gomez, Pan, Hsieh, Sánchez, Bizyaev, Almeida, Thilker, Lotz,
  Jones, Nair, Andrews, \& Drory}]{hwang_anomalously_2019}
Hwang, H.-C., Barrera-Ballesteros, J.~K., Heckman, T.~M., {et~al.} 2019, The
  Astrophysical Journal, 872, 144, \dodoi{10.3847/1538-4357/aaf7a3}

\bibitem[{{James} {et~al.}(2020){James}, {Kumari}, {Emerick}, {Koposov},
  {McQuinn}, {Stark}, {Belokurov}, \& {Maiolino}}]{James_2020}
{James}, B.~L., {Kumari}, N., {Emerick}, A., {et~al.} 2020, \mnras, 495, 2564,
  \dodoi{10.1093/mnras/staa1280}

\bibitem[{{James} {et~al.}(2013){James}, {Tsamis}, {Walsh}, {Barlow}, \&
  {Westmoquette}}]{James_2013}
{James}, B.~L., {Tsamis}, Y.~G., {Walsh}, J.~R., {Barlow}, M.~J., \&
  {Westmoquette}, M.~S. 2013, \mnras, 430, 2097, \dodoi{10.1093/mnras/stt034}

\bibitem[{Kennicutt(1998)}]{Kennicutt_Jr__1998}
Kennicutt, J. 1998, The Astrophysical Journal, 498, 541, \dodoi{10.1086/305588}

\bibitem[{Kewley {et~al.}(2019)Kewley, Nicholls, \& Sutherland}]{kewley_2019}
Kewley, L.~J., Nicholls, D.~C., \& Sutherland, R.~S. 2019, Annual Review of
  Astronomy and Astrophysics, 57, 511,
  \dodoi{10.1146/annurev-astro-081817-051832}

\bibitem[{Kewley {et~al.}(2010)Kewley, Rupke, Jabran~Zahid, Geller, \&
  Barton}]{kewley_metallicity_2010}
Kewley, L.~J., Rupke, D., Jabran~Zahid, H., Geller, M.~J., \& Barton, E.~J.
  2010, The Astrophysical Journal, 721, L48,
  \dodoi{10.1088/2041-8205/721/1/L48}

\bibitem[{Kisielius {et~al.}(2009)Kisielius, Storey, Ferland, \&
  Keenan}]{kisielius_electron-impact_2009}
Kisielius, R., Storey, P.~J., Ferland, G.~J., \& Keenan, F.~P. 2009, Monthly
  Notices of the Royal Astronomical Society, 397, 903,
  \dodoi{10.1111/j.1365-2966.2009.14989.x}

\bibitem[{{Kobulnicky} \& {Skillman}(1997)}]{KS_1997}
{Kobulnicky}, H.~A., \& {Skillman}, E.~D. 1997, \apj, 489, 636,
  \dodoi{10.1086/304830}

\bibitem[{{Kroupa}(2001)}]{Kroupa_2001}
{Kroupa}, P. 2001, \mnras, 322, 231, \dodoi{10.1046/j.1365-8711.2001.04022.x}

\bibitem[{{Kumari} {et~al.}(2019){Kumari}, {James}, {Irwin}, \&
  {Aloisi}}]{Kumari_2019}
{Kumari}, N., {James}, B.~L., {Irwin}, M.~J., \& {Aloisi}, A. 2019, \mnras,
  485, 1103, \dodoi{10.1093/mnras/stz343}

\bibitem[{{Law} {et~al.}(2012){Law}, {Steidel}, {Shapley}, {Nagy}, {Reddy}, \&
  {Erb}}]{Law_2012}
{Law}, D.~R., {Steidel}, C.~C., {Shapley}, A.~E., {et~al.} 2012, \apj, 745, 85,
  \dodoi{10.1088/0004-637X/745/1/85}

\bibitem[{{Lee} \& {Skillman}(2004)}]{LS_2004}
{Lee}, H., \& {Skillman}, E.~D. 2004, \apj, 614, 698, \dodoi{10.1086/423735}

\bibitem[{Lee {et~al.}(2006)Lee, Skillman, Cannon, Jackson, Gehrz, Polomski, \&
  Woodward}]{lee_extending_2006}
Lee, H., Skillman, E.~D., Cannon, J.~M., {et~al.} 2006, The Astrophysical
  Journal, 647, 970, \dodoi{10.1086/505573}

\bibitem[{{Lehnert} \& {Heckman}(1996)}]{Lehnert_1996}
{Lehnert}, M.~D., \& {Heckman}, T.~M. 1996, \apj, 462, 651,
  \dodoi{10.1086/177180}

\bibitem[{{Leitherer} {et~al.}(1999){Leitherer}, {Schaerer}, {Goldader},
  {Delgado}, {Robert}, {Kune}, {de Mello}, {Devost}, \&
  {Heckman}}]{starburst99}
{Leitherer}, C., {Schaerer}, D., {Goldader}, J.~D., {et~al.} 1999, \apjs, 123,
  3, \dodoi{10.1086/313233}

\bibitem[{L{\'{o}}pez-S{\'{a}}nchez {et~al.}(2012)L{\'{o}}pez-S{\'{a}}nchez,
  Dopita, Kewley, Zahid, Nicholls, \&
  Scharwächter}]{lopez-sanchez_eliminating_2012}
L{\'{o}}pez-S{\'{a}}nchez, {\'{A}}.~R., Dopita, M.~A., Kewley, L.~J., {et~al.}
  2012, Monthly Notices of the Royal Astronomical Society, 426, 2630,
  \dodoi{10.1111/j.1365-2966.2012.21145.x}

\bibitem[{Luo {et~al.}(2021)Luo, Heckman, Hwang, Rowlands, Sánchez-Menguiano,
  Riffel, Bizyaev, Andrews, Fernández-Trincado, Drory, Almeida, Maiolino,
  Lane, \& Argudo-Fernández}]{luo_evidence_2021}
Luo, Y., Heckman, T., Hwang, H.-C., {et~al.} 2021, The Astrophysical Journal,
  908, 183, \dodoi{10.3847/1538-4357/abd1df}

\bibitem[{{Luridiana} {et~al.}(2013){Luridiana}, {Morisset}, \& {Shaw}}]{pyneb}
{Luridiana}, V., {Morisset}, C., \& {Shaw}, R.~A. 2013, {PyNeb: Analysis of
  emission lines}, Astrophysics Source Code Library, record ascl:1304.021.
\newblock \doeprint{1304.021}

\bibitem[{{Luridiana} {et~al.}(2015){Luridiana}, {Morisset}, \&
  {Shaw}}]{pyneb_2015}
---. 2015, \aap, 573, A42, \dodoi{10.1051/0004-6361/201323152}

\bibitem[{Ly {et~al.}(2016)Ly, Malkan, Rigby, \& Nagao}]{Ly_2016}
Ly, C., Malkan, M.~A., Rigby, J.~R., \& Nagao, T. 2016, The Astrophysical
  Journal, 828, 67, \dodoi{10.3847/0004-637x/828/2/67}

\bibitem[{Ma {et~al.}(2017)Ma, Hopkins, Feldmann, Torrey, Faucher-Giguère, \&
  Kereš}]{ma_why_2017}
Ma, X., Hopkins, P.~F., Feldmann, R., {et~al.} 2017, Monthly Notices of the
  Royal Astronomical Society, stx034, \dodoi{10.1093/mnras/stx034}

\bibitem[{{Maiolino} \& {Mannucci}(2019)}]{MM_2019}
{Maiolino}, R., \& {Mannucci}, F. 2019, \aapr, 27, 3,
  \dodoi{10.1007/s00159-018-0112-2}

\bibitem[{Mannucci {et~al.}(2010)Mannucci, Cresci, Maiolino, Marconi, \&
  Gnerucci}]{mannucci_fundamental_2010}
Mannucci, F., Cresci, G., Maiolino, R., Marconi, A., \& Gnerucci, A. 2010,
  Monthly Notices of the Royal Astronomical Society, 408, 2115,
  \dodoi{10.1111/j.1365-2966.2010.17291.x}

\bibitem[{{Martin} {et~al.}(2002){Martin}, {Kobulnicky}, \&
  {Heckman}}]{Martin_2002}
{Martin}, C.~L., {Kobulnicky}, H.~A., \& {Heckman}, T.~M. 2002, \apj, 574, 663,
  \dodoi{10.1086/341092}

\bibitem[{Masters(2005)}]{masters_2005}
Masters, K.~L. 2005, PhD thesis, Cornell University, New York, USA

\bibitem[{{Moll{\'a}} \& {D{\'\i}az}(2005)}]{Molla_2005}
{Moll{\'a}}, M., \& {D{\'\i}az}, A.~I. 2005, \mnras, 358, 521,
  \dodoi{10.1111/j.1365-2966.2005.08782.x}

\bibitem[{Morrissey {et~al.}(2018)Morrissey, Matuszewski, Martin, Neill, Epps,
  Fucik, Weber, Darvish, Adkins, Allen, Bartos, Belicki, Cabak, Callahan,
  Cowley, Crabill, Deich, Delecroix, Doppman, Hilyard, James, Kaye, Kokorowski,
  Kwok, Lanclos, Milner, Moore, O’Sullivan, Parihar, Park, Phillips, Rizzi,
  Rockosi, Rodriguez, Salaun, Seaman, Sheikh, Weiss, \&
  Zarzaca}]{morrissey_keck_2018}
Morrissey, P., Matuszewski, M., Martin, D.~C., {et~al.} 2018, The Astrophysical
  Journal, 864, 93, \dodoi{10.3847/1538-4357/aad597}

\bibitem[{Moster {et~al.}(2012)Moster, Naab, \& White}]{Moster_2012}
Moster, B.~P., Naab, T., \& White, S. D.~M. 2012, Monthly Notices of the Royal
  Astronomical Society, 428, 3121, \dodoi{10.1093/mnras/sts261}

\bibitem[{Moustakas {et~al.}(2013)Moustakas, Coil, Aird, Blanton, Cool,
  Eisenstein, Mendez, Wong, Zhu, \& Arnouts}]{Moustakas_2013}
Moustakas, J., Coil, A.~L., Aird, J., {et~al.} 2013, The Astrophysical Journal,
  767, 50, \dodoi{10.1088/0004-637x/767/1/50}

\bibitem[{{Newville} {et~al.}(2021){Newville}, {Otten}, {Nelson}, {Ingargiola},
  {Stensitzki}, {Allan}, {Fox}, {Carter}, {Micha{\l}}, {Osborn}, {Pustakhod},
  {Lneuhaus}, {Weigand}, {Glenn}, {Deil}, {Mark}, {Hansen}, {Pasquevich},
  {Foks}, {Zobrist}, {Frost}, {Beelen}, {Stuermer}, {Azelcer}, {Hannum},
  {Polloreno}, {Hedegaard Nielsen}, {Caldwell}, {Almarza}, \&
  {Persaud}}]{lmfit}
{Newville}, M., {Otten}, R., {Nelson}, A., {et~al.} 2021, {lmfit/lmfit-py:
  1.0.3}, 1.0.3, Zenodo,  Zenodo, \dodoi{10.5281/zenodo.5570790}

\bibitem[{Olivier {et~al.}(2022)Olivier, Berg, Chisholm, Erb, Pogge, \&
  Skillman}]{Olivier_2022}
Olivier, G.~M., Berg, D.~A., Chisholm, J., {et~al.} 2022, The Astrophysical
  Journal, 938, 16, \dodoi{10.3847/1538-4357/ac8f2c}

\bibitem[{{O'Sullivan} \& {Chen}(2020)}]{OC_2020}
{O'Sullivan}, D., \& {Chen}, Y. 2020, arXiv e-prints, arXiv:2011.05444,
  \dodoi{10.48550/arXiv.2011.05444}

\bibitem[{O'Sullivan {et~al.}(2020)O'Sullivan, Martin, Matuszewski, Hoadley,
  Hamden, Neill, Lin, \& Parihar}]{osullivan_flashes_2020}
O'Sullivan, D.~B., Martin, C., Matuszewski, M., {et~al.} 2020, The
  Astrophysical Journal, 894, 3, \dodoi{10.3847/1538-4357/ab838c}

\bibitem[{P{\'{e}}rez-Montero(2017)}]{P_rez_Montero_2017}
P{\'{e}}rez-Montero, E. 2017, Publications of the Astronomical Society of the
  Pacific, 129, 043001, \dodoi{10.1088/1538-3873/aa5abb}

\bibitem[{{Petrosian}(1976)}]{Petrosian_1976}
{Petrosian}, V. 1976, \apjl, 210, L53, \dodoi{10.1086/182301}

\bibitem[{Pilkington {et~al.}(2012)Pilkington, Few, Gibson, Calura,
  Michel-Dansac, Thacker, Mollá, Matteucci, Rahimi, Kawata, Kobayashi, Brook,
  Stinson, Couchman, Bailin, \& Wadsley}]{pilkington_metallicity_2012}
Pilkington, K., Few, C.~G., Gibson, B.~K., {et~al.} 2012, \aap, 540, A56,
  \dodoi{10.1051/0004-6361/201117466}

\bibitem[{Rich {et~al.}(2012)Rich, Torrey, Kewley, Dopita, \&
  Rupke}]{rich_integral_2012}
Rich, J.~A., Torrey, P., Kewley, L.~J., Dopita, M.~A., \& Rupke, D. S.~N. 2012,
  \apj, 753, 5, \dodoi{10.1088/0004-637X/753/1/5}

\bibitem[{Rupke {et~al.}(2010{\natexlab{a}})Rupke, Kewley, \&
  Barnes}]{rupke_galaxy_2010}
Rupke, D. S.~N., Kewley, L.~J., \& Barnes, J.~E. 2010{\natexlab{a}}, \apj, 710,
  L156, \dodoi{10.1088/2041-8205/710/2/L156}

\bibitem[{Rupke {et~al.}(2010{\natexlab{b}})Rupke, Kewley, \&
  Chien}]{rupke_gas-phase_2010}
Rupke, D. S.~N., Kewley, L.~J., \& Chien, L.-H. 2010{\natexlab{b}}, \apj, 723,
  1255, \dodoi{10.1088/0004-637X/723/2/1255}

\bibitem[{{S{\'a}nchez-Monge} {et~al.}(2018){S{\'a}nchez-Monge}, {Schilke},
  {Ginsburg}, {Cesaroni}, \& {Schmiedeke}}]{cont_2018}
{S{\'a}nchez-Monge}, {\'A}., {Schilke}, P., {Ginsburg}, A., {Cesaroni}, R., \&
  {Schmiedeke}, A. 2018, \aap, 609, A101, \dodoi{10.1051/0004-6361/201730425}

\bibitem[{{Sanders} {et~al.}(2021){Sanders}, {Shapley}, {Jones}, {Reddy},
  {Kriek}, {Siana}, {Coil}, {Mobasher}, {Shivaei}, {Dav{\'e}}, {Azadi},
  {Price}, {Leung}, {Freeman}, {Fetherolf}, {de Groot}, {Zick}, \&
  {Barro}}]{Sanders_2021}
{Sanders}, R.~L., {Shapley}, A.~E., {Jones}, T., {et~al.} 2021, \apj, 914, 19,
  \dodoi{10.3847/1538-4357/abf4c1}

\bibitem[{Schlafly \& Finkbeiner(2011)}]{schlafly_measuring_2011}
Schlafly, E.~F., \& Finkbeiner, D.~P. 2011, The Astrophysical Journal, 737,
  103, \dodoi{10.1088/0004-637X/737/2/103}

\bibitem[{Sharda {et~al.}(2021)Sharda, Krumholz, Wisnioski, Acharyya,
  Federrath, \& Forbes}]{sharda2021origin}
Sharda, P., Krumholz, M.~R., Wisnioski, E., {et~al.} 2021, Monthly Notices of
  the Royal Astronomical Society, 504, 53, \dodoi{10.1093/mnras/stab868}

\bibitem[{{Sharda} {et~al.}(2021){Sharda}, {Krumholz}, {Wisnioski}, {Forbes},
  {Federrath}, \& {Acharyya}}]{Sharda_Gradient_2021}
{Sharda}, P., {Krumholz}, M.~R., {Wisnioski}, E., {et~al.} 2021, \mnras, 502,
  5935, \dodoi{10.1093/mnras/stab252}

\bibitem[{{Stasi\'{n}ska, G.} {et~al.}(2007){Stasi\'{n}ska, G.},
  {Tenorio-Tagle, G.}, {Rodr\'{\i}guez, M.}, \& {Henney, W. J.}}]{ST07}
{Stasi\'{n}ska, G.}, {Tenorio-Tagle, G.}, {Rodr\'{\i}guez, M.}, \& {Henney, W.
  J.} 2007, A\&A, 471, 193, \dodoi{10.1051/0004-6361:20065675}

\bibitem[{Storey \& Hummer(1995)}]{storey_recombination_1995}
Storey, P.~J., \& Hummer, D.~G. 1995, Monthly Notices of the Royal Astronomical
  Society, 272, 41, \dodoi{10.1093/mnras/272.1.41}

\bibitem[{Storey {et~al.}(2014)Storey, Sochi, \& Badnell}]{Storey_2014}
Storey, P.~J., Sochi, T., \& Badnell, N.~R. 2014, Monthly Notices of the Royal
  Astronomical Society, 441, 3028, \dodoi{10.1093/mnras/stu777}

\bibitem[{{Storey} \& {Zeippen}(2000)}]{SZ_2000}
{Storey}, P.~J., \& {Zeippen}, C.~J. 2000, \mnras, 312, 813,
  \dodoi{10.1046/j.1365-8711.2000.03184.x}

\bibitem[{Tenorio-Tagle(1996)}]{tenorio-tagle_interstellar_1996}
Tenorio-Tagle, G. 1996, The Astronomical Journal, 111, 1641,
  \dodoi{10.1086/117903}

\bibitem[{Tissera {et~al.}(2019)Tissera, Rosas-Guevara, Bower, Crain, Lagos,
  Schaller, Schaye, \& Theuns}]{tissera_oxygen_2019}
Tissera, P.~B., Rosas-Guevara, Y., Bower, R.~G., {et~al.} 2019, Monthly Notices
  of the Royal Astronomical Society, 482, 2208, \dodoi{10.1093/mnras/sty2817}

\bibitem[{Torrey {et~al.}(2012)Torrey, Cox, Kewley, \&
  Hernquist}]{torrey_metallicity_2012}
Torrey, P., Cox, T.~J., Kewley, L., \& Hernquist, L. 2012, \apj, 746, 108,
  \dodoi{10.1088/0004-637X/746/1/108}

\bibitem[{Tsatsi {et~al.}(2017)Tsatsi, Lyubenova, van~de Ven, Chang, Aguerri,
  Falcón-Barroso, \& Macciò}]{tsatsi_califa_2017}
Tsatsi, A., Lyubenova, M., van~de Ven, G., {et~al.} 2017, \aap, 606, A62,
  \dodoi{10.1051/0004-6361/201630218}

\bibitem[{{Veilleux} {et~al.}(2020){Veilleux}, {Maiolino}, {Bolatto}, \&
  {Aalto}}]{Veilleux_2020}
{Veilleux}, S., {Maiolino}, R., {Bolatto}, A.~D., \& {Aalto}, S. 2020, \aapr,
  28, 2, \dodoi{10.1007/s00159-019-0121-9}

\bibitem[{{Vincenzo} {et~al.}(2016){Vincenzo}, {Matteucci}, {Belfiore}, \&
  {Maiolino}}]{Vincenzo_2016}
{Vincenzo}, F., {Matteucci}, F., {Belfiore}, F., \& {Maiolino}, R. 2016,
  \mnras, 455, 4183, \dodoi{10.1093/mnras/stv2598}

\bibitem[{Wang {et~al.}(2019)Wang, Jones, Treu, Hirtenstein, Brammer, Daddi,
  Meng, Morishita, Abramson, Henry, Peng, Schmidt, Sharon, Trenti, \&
  Vulcani}]{wang_discovery_2019}
Wang, X., Jones, T.~A., Treu, T., {et~al.} 2019, The Astrophysical Journal,
  882, 94, \dodoi{10.3847/1538-4357/ab3861}

\bibitem[{{Wheeler} {et~al.}(2019){Wheeler}, {Hopkins}, {Pace},
  {Garrison-Kimmel}, {Boylan-Kolchin}, {Wetzel}, {Bullock}, {Kere{\v{s}}},
  {Faucher-Gigu{\`e}re}, \& {Quataert}}]{Wheeler_2019}
{Wheeler}, C., {Hopkins}, P.~F., {Pace}, A.~B., {et~al.} 2019, \mnras, 490,
  4447, \dodoi{10.1093/mnras/stz2887}

\bibitem[{Whitaker {et~al.}(2014)Whitaker, Franx, Leja, van Dokkum, Henry,
  Skelton, Fumagalli, Momcheva, Brammer, Labbé, Nelson, \&
  Rigby}]{whitaker_constraining_2014}
Whitaker, K.~E., Franx, M., Leja, J., {et~al.} 2014, The Astrophysical Journal,
  795, 104, \dodoi{10.1088/0004-637X/795/2/104}

\bibitem[{Whitmore {et~al.}(2010)Whitmore, Chandar, Schweizer, Rothberg,
  Leitherer, Rieke, Rieke, Blair, Mengel, \& Alonso-Herrero}]{Whitmore_2010}
Whitmore, B.~C., Chandar, R., Schweizer, F., {et~al.} 2010, The Astronomical
  Journal, 140, 75, \dodoi{10.1088/0004-6256/140/1/75}

\bibitem[{Willick {et~al.}(1997)Willick, Courteau, Faber, Burstein, Dekel, \&
  Strauss}]{willick_homogeneous_1997}
Willick, J.~A., Courteau, S., Faber, S.~M., {et~al.} 1997, The Astrophysical
  Journal Supplement Series, 109, 333, \dodoi{10.1086/312983}

\bibitem[{Xu {et~al.}(2022)Xu, Heckman, Henry, Berg, Chisholm, James, Martin,
  Stark, Aloisi, Amor{\'{\i} }n, Arellano-C{\'{o}}rdova, Bordoloi, Charlot,
  Chen, Hayes, Mingozzi, Sugahara, Kewley, Ouchi, Scarlata, \&
  Steidel}]{xu_classy_2022}
Xu, X., Heckman, T., Henry, A., {et~al.} 2022, The Astrophysical Journal, 933,
  222, \dodoi{10.3847/1538-4357/ac6d56}

\bibitem[{{Yasuda} {et~al.}(2001){Yasuda}, {Fukugita}, {Narayanan}, {Lupton},
  {Strateva}, {Strauss}, {Ivezi{\'c}}, {Kim}, {Hogg}, {Weinberg}, {Shimasaku},
  {Loveday}, {Annis}, {Bahcall}, {Blanton}, {Brinkmann}, {Brunner}, {Connolly},
  {Csabai}, {Doi}, {Hamabe}, {Ichikawa}, {Ichikawa}, {Johnston}, {Knapp},
  {Kunszt}, {Lamb}, {McKay}, {Munn}, {Nichol}, {Okamura}, {Schneider},
  {Szokoly}, {Vogeley}, {Watanabe}, \& {York}}]{Yasuda_01}
{Yasuda}, N., {Fukugita}, M., {Narayanan}, V.~K., {et~al.} 2001, \aj, 122,
  1104, \dodoi{10.1086/322093}

\bibitem[{Yates {et~al.}(2020)Yates, Schady, Chen, Schweyer, \&
  Wiseman}]{yates_present-day_2020}
Yates, R.~M., Schady, P., Chen, T.-W., Schweyer, T., \& Wiseman, P. 2020, \aap,
  634, A107, \dodoi{10.1051/0004-6361/201936506}

\bibitem[{{Zahid} {et~al.}(2014){Zahid}, {Kashino}, {Silverman}, {Kewley},
  {Daddi}, {Renzini}, {Rodighiero}, {Nagao}, {Arimoto}, {Sanders},
  {Kartaltepe}, {Lilly}, {Maier}, {Geller}, {Capak}, {Carollo}, {Chu},
  {Hasinger}, {Ilbert}, {Kajisawa}, {Koekemoer}, {Kovacs}, {Le F{\`e}vre},
  {Masters}, {McCracken}, {Onodera}, {Scoville}, {Strazzullo}, {Sugiyama},
  {Taniguchi}, \& {COSMOS Team}}]{Zahid_2014_dust}
{Zahid}, H.~J., {Kashino}, D., {Silverman}, J.~D., {et~al.} 2014, \apj, 792,
  75, \dodoi{10.1088/0004-637X/792/1/75}

\end{thebibliography}
\bibliographystyle{aasjournal}



\end{document}